\documentclass[reprint,superscriptaddress,amsmath,aps,floatfix,prb,showpacs,showkeys]{revtex4-1}
\usepackage{graphicx}
\usepackage{dcolumn}
\usepackage{bm}
\usepackage[version=3]{mhchem}
\usepackage{multirow}
\usepackage[protrusion=true,expansion=true,final]{microtype}
\usepackage[bitstream-charter]{mathdesign}
\usepackage[T1]{fontenc}
\usepackage{booktabs}
\usepackage[usenames,dvipsnames]{xcolor}
\usepackage[bookmarks=false,colorlinks]{hyperref}
\hypersetup{
	linkcolor= Black,        
	citecolor = Black,        
	filecolor=Black,        
	urlcolor=Black,         
	pdfauthor={Debangshu Mukherjee,Sergei Prokhorenko, Leixin Miao, Ke Wang, Eric Bousquet, Venkatraman Gopalan and Nasim Alem},
	pdftitle={Atomic Scale Measurement of Polar Entropy},
}
\usepackage{xfrac}

\begin{document}
	\title{Atomic Scale Measurement of Polar Entropy }
	\author{Debangshu\ Mukherjee}
	\affiliation{Department of Materials Science and Engineering, The Pennsylvania State University, University Park, Pennsylvania 16802, USA}
	\author{Sergei\ Prokhorenko}
	\affiliation{Theoretical Materials Physics Q-MAT CESAM, University of Li\`{e}ge, Sart Tilman B-4000, Belgium}
	\affiliation{Physics Department and Institute for Nanoscience and Engineering, University of Arkansas, Fayetteville, Arkansas 72701, USA}
	\author{Leixin\ Miao}
	\affiliation{Department of Materials Science and Engineering, The Pennsylvania State University, University Park, Pennsylvania 16802, USA}
	\author{Ke\ Wang}
	\affiliation{Materials Characterization Laboratory, Materials Research Institute, The Pennsylvania State University, University Park, Pennsylvania 16802, USA}
	\author{Eric\ Bousquet}
	\affiliation{Theoretical Materials Physics Q-MAT CESAM, University of Li\`{e}ge, Sart Tilman B-4000, Belgium}
	\author{Venkatraman\ Gopalan}
	\affiliation{Department of Materials Science and Engineering, The Pennsylvania State University, University Park, Pennsylvania 16802, USA}
	\affiliation{Materials Characterization Laboratory, Materials Research Institute, The Pennsylvania State University, University Park, Pennsylvania 16802, USA}
	\author{Nasim\ Alem}
	\email{alem@matse.psu.edu}
	\affiliation{Department of Materials Science and Engineering, The Pennsylvania State University, University Park, Pennsylvania 16802, USA}
	\affiliation{Materials Characterization Laboratory, Materials Research Institute, The Pennsylvania State University, University Park, Pennsylvania 16802, USA}
	
	\begin{abstract}
		Entropy is a fundamental thermodynamic quantity that is a measure of the accessible microstates available to a system\cite{GibbsMain}, with the stability of a system determined by the magnitude of the total entropy of the system. This is valid across truly mind boggling length scales - from nanoparticles to galaxies\cite{heatDeath,entropy_nano}. However, quantitative measurements of entropy change using calorimetry are predominantly macroscopic, with direct atomic scale measurements being exceedingly rare\cite{EntropyMetrology}.  Here for the first time, we experimentally quantify the polar configurational entropy (in meV/K) using sub-\r{a}ngstr\"{o}m resolution aberration corrected scanning transmission electron microscopy.  This is performed in a single crystal of the prototypical ferroelectric \ce{LiNbO3} through the quantification of the niobium and oxygen atom column deviations from their paraelectric positions. Significant excursions of the niobium--oxygen polar displacement away from its symmetry constrained direction is seen in single domain regions which increases in the proximity of domain walls. Combined with first principles theory plus mean field effective Hamiltonian methods, we demonstrate the variability in the polar order parameter, which is stabilized by an increase in the magnitude of the configurational entropy. This study presents a powerful tool to quantify entropy from atomic displacements and demonstrates its dominant role in local symmetry breaking at finite temperatures in classic, nominally Ising ferroelectrics.
	\end{abstract}
	
	\pacs{65.40.gd, 68.37.Ma, 77.80.Dj, 77.84.Ek}
	\keywords{\ce{LiNbO3}, Ferroelectricity, Domain Wall, Polar Entropy, Aberration Corrected STEM}
	\maketitle

\section{\label{sec:Intro}Introduction}

While absolute entropy, a fundamental thermodynamic parameter, is difficult to experimentally measure macroscopically, a change in entropy $\mathrm{\left( \Delta S = \sfrac{\Delta Q_{rev}}{T} \right)}$, is usually measured using calorimetry, where $\mathrm{\Delta Q_{rev}}$ is the reversible heat supplied to the system at a constant temperature T\cite{GibbsMain,EntropyMetrology}. At absolute zero $\mathrm{\left( T = 0K \right)}$, the total entropy of a perfect crystal free of dopants is zero.  Upon addition of reversible heat to the system, the entropy increases. Directly measuring the absolute entropy of the system through characterizing the microscopic configurations, or the microstates, is challenging, since it increases exponentially with the number of available microstates.  Such enormously large numbers of microstates are also involved in condensed matter systems where a dopant atom may choose any one of equivalent atomic sites in a periodic lattice. The perturbation in atom positions from the crystal sites thus leads to an increase in the configurational entropy - which can be quantified through the probability distributions of the perturbations. Such configurational entropy may arise for example in ferroelectric crystals - due to perturbations in the order parameter. The order parameter of a ferroelectric system is the spontaneous polarization, which arises as a consequence of the  polar displacements\cite{modern_ferro}. The advances in aberration corrected scanning transmission electron microscopy(STEM) - have now made it possible to quantify displacements with a precision approaching a single picometer\cite{nelsonBFO,pico1,CTO_Domain}, and even below a single picometer\cite{pico_catalyst}. Recent results have demonstrated the feasibility of visualizing 2pm magnitude charge density waves even at cryogenic temperatures with dark field STEM\cite{Lena_CDW}. We build upon these advances to perform picometer precison quantification of polar displacements to quantify the variation in the order parameter in the well-known optical ferroelectric \ce{LiNbO3} relative to its' ideal ferroelectric structure. Throughout the rest of this work, we will refer to this entropy arising from the variability in polar displacements as polar entropy.

\section{\label{sec:MeasurePolar}Measurement of Niobium--Oxygen polar displacements}
Ferroelectric materials have a spontaneous and switchable electrical polarization, which is a consequence of the lattice distortions in the crystal structure that break inversion symmetry\cite{lines_glass}. Regions of uniform polarization are called domains, with the boundary between two adjacent domains referred to as a domain wall\cite{DomainBook,DomainWallReview}. Since the ferroelectric polarization is a consequence of crystal lattice distortions, the possible polar vectors can occur only along certain symmetry allowed crystallographic directions. As a uniaxial displacive ferroelectric (space group R3c), the origin of the spontaneous polarization in \ce{LiNbO3} is a consequence of the niobium and lithium cation displacements with respect to the oxygen octahedral center along either the $\left(0001\right)$ or the $\left(000\bar{1}\right)$ crystallographic axes, and thus the polarization vectors are restricted to only $\left\langle 0001 \right\rangle$ direction (also labeled as z- or 3- direction)\cite{LNO_review}.  Classical uniaxial ferroelectrics such as \ce{LiNbO3} have been long thought of as Ising like where the polarization is only along the two symmetry restricted directions, which transitions to zero at the domain wall, since lattice distortions away from the symmetry restricted polarization directions have a high energy cost associated with them\cite{ChargedDW1,ChargedDW2}.  However, recent research have pointed out that fluctuations away from the Ising polarization direction do exist in other ferroelectrics - most notably \ce{PbTiO3}, with Bloch and N{\'e}el components arising at domain walls \cite{BlochNeelIsing,BlochPTO,NeelWalls,IniguezPTO}. However, such deviations, as per the authors' knowldge have never been observed before in \ce{LiNbO3}.

\begin{figure*}
\centering
\includegraphics[width=\textwidth]{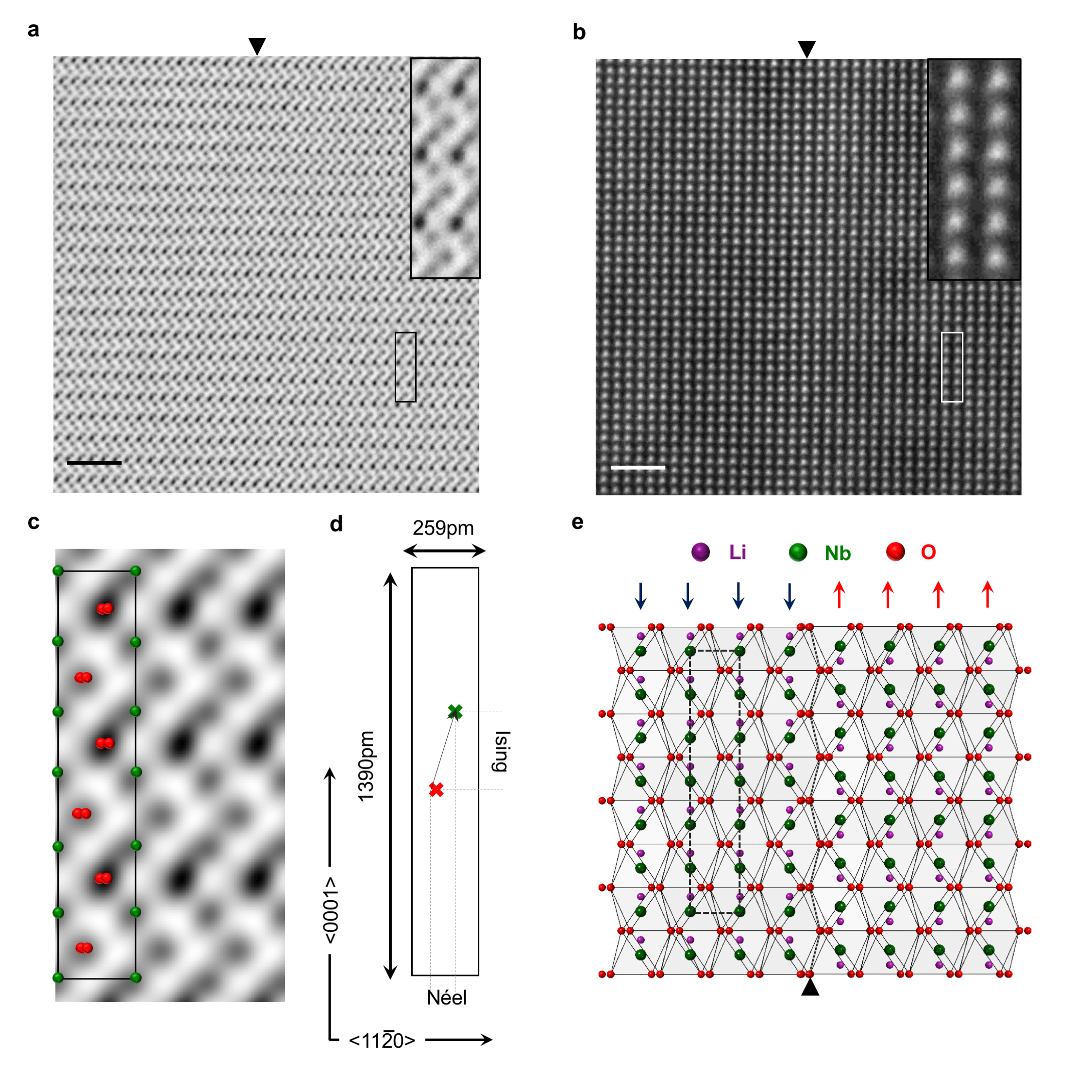}
\caption{\label{fig:schematic} Schematic of electron microscopy experiments. (a) Aberration-corrected Bright Field (BF)-STEM image of a domain wall in \ce{LiNbO3}, with the wall location marked by the black triangle, with a zoomed section in inset showing oxygen and niobium positions. Scale bar is 2nm.  (b) Simultaneously collected Annular Dark Field (ADF)-STEM image of the region imaged in \autoref{fig:schematic}(a), with the wall location marked by the black triangle. The zoomed section in inset shows the niobium atoms. Scale bar is 2nm. (c) Averaged mean unit cell from the experimental datasets with the niobium positions in green and the oxygen atoms in red. The unit cell is shown as the black rectangle. (d) The schematic of the unit cell with the experimentally measured long and short dimensions. The niobium and oxygen centers are shown as green and red crosses with the projected Ising and N{\'e}el displacement directions. (e) Atomic model of \ce{LiNbO3} crystal structure viewed from the $\mathrm{\left[ 1 \bar{1}00 \right]}$ zone axis, with lithium atoms in purple, niobium atoms in green and oxygen atoms in red. The average unit cell for polarization calculations is shown as a black dashed box with the arrows referring to the polarization direction.}
\end{figure*}

To visualize the atom positions at the $180^{\circ}$ domain wall and also at the bulk domain, we imaged the electron transparent \ce{LiNbO3} sample from the $\mathrm{\left[ 1\bar{1}00 \right] }$  crystallographic zone axis so that the Ising displacements lie in plane. While both bright field (BF) and annular dark field (ADF) STEM images were acquired (\autoref{fig:schematic}(a) and \autoref{fig:schematic}(b)), we exclusively use BF-STEM images for the quantification of polar displacements since both the niobium and oxygen atom positions and their relative displacements can be quantified. This technique has been previously demonstrated as a viable pathway for the determination of the cation and oxygen atom positions simultaneously\cite{multivariateYMO}, and is less susceptible to specimen tilt and defocus in comparison to annular bright field (ABF)-STEM\cite{albinaBFSTEM1,albinaBFSTEM2}. The samples imaged in this experiment were approximately 25nm thick, as determined from electron energy loss spectroscopy (EELS) inelastic mean free path measurements (see \autoref{fig:thickness} in appendix)\cite{eels_thickness}.

The total polar displacements are calculated per unit cell, with respect to a mean unit cell calculated from the entire image (\autoref{fig:schematic}(c)). The mean unit cell calculated from the STEM experimental data has the dimensions of $\mathrm{1390pm \times 259pm}$ -- which is within 2\% of the simulated \ce{LiNbO3} \emph{R3c} unit cell parameters of $\mathrm{1412.92pm \times 261.15pm }$ when viewed from the $\mathrm{\left[ 1\bar{1}00 \right] }$  crystallographic zone axis\cite{materials_project}. As demonstrated in \autoref{fig:schematic}(d), displacements along $\mathrm{\left\langle 0001 \right\rangle }$ are the Ising displacements, while those along $\mathrm{\left\langle 11\bar{2}0 \right\rangle }$ are the N{\'e}el displacements. To determine the polar displacements, we first assigned the measured atom positions to their corresponding \ce{LiNbO3} unit cell positions, and then generated an average unit cell by summing all the individual unit cells throughout the BF-STEM image (\autoref{fig:schematic}(d)).  The oxygen and niobium centers of mass for each individual unit cell (\autoref{fig:schematic}(e)) were subsequently compared to the center of the calculated mean unit cell to determine the displacement vectors for both niobium and oxygen atoms for each \ce{LiNbO3} unit cell imaged. The Nb-O polar displacement were then measured as a vector subtraction of the oxygen displacement vector from the niobium displacement vector. This calculated polar displacement vector was subsequently decomposed into its corresponding Ising and N{\'e}el components along the $\mathrm{\left\langle 0001 \right\rangle}$ and $\mathrm{\left\langle 11\bar{2}0 \right\rangle}$ directions respectively to obtain the individual polar components.

\begin{figure*}
\centering
\includegraphics[width=\textwidth]{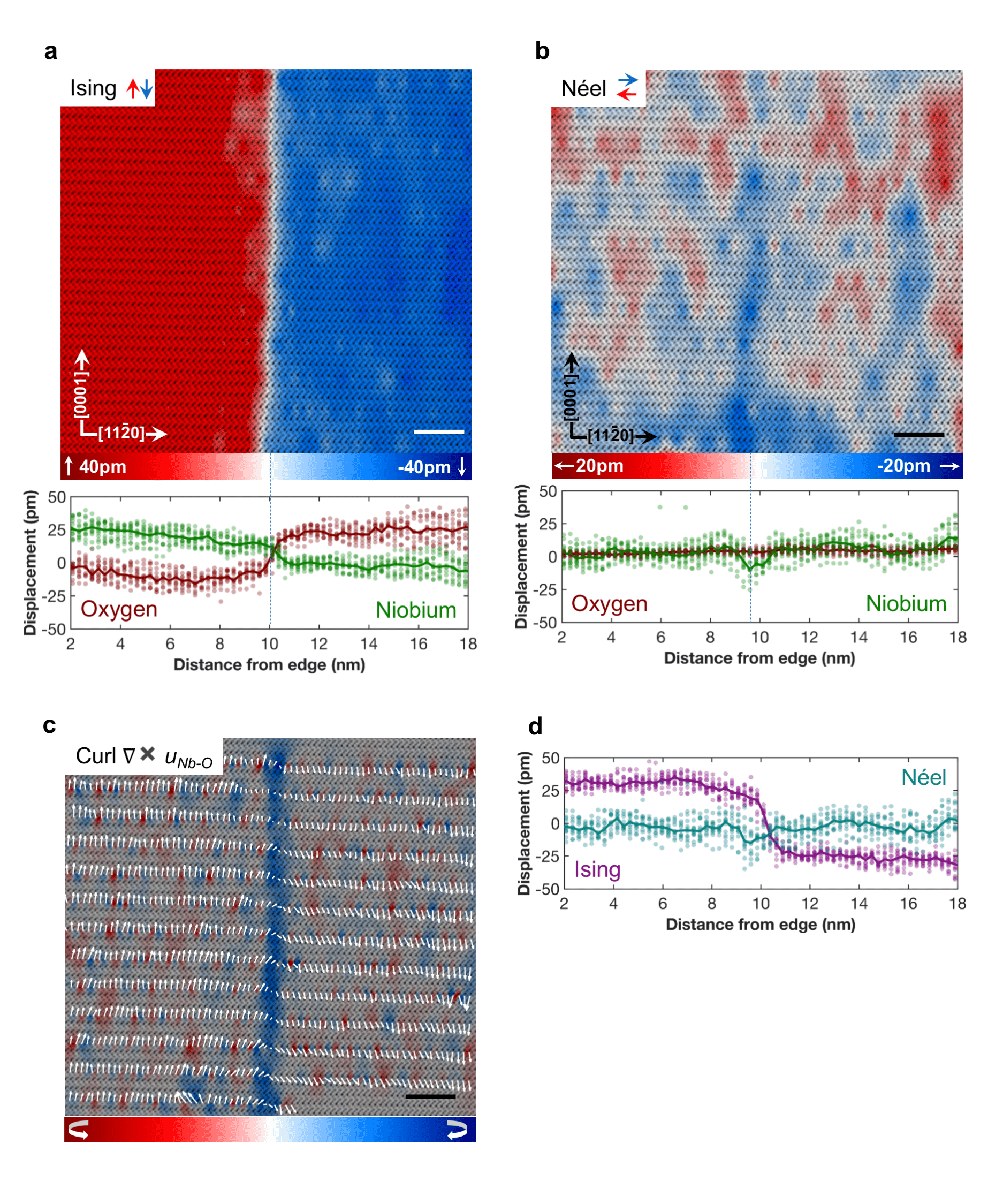}
\caption{\label{fig:polarMap} Polar displacements measured with BF-STEM. (a) \ce{LiNbO3} domain wall imaged from $\mathrm{\left[ 1\bar{1}00 \right] }$ zone axis with the polar Ising niobium-oxygen displacements overlaid. Scale bar is 2 nm. The Ising niobium and oxygen along the $\mathrm{\left\langle 0001 \right\rangle }$ direction in green and red respectively with the solid lines referring to the averages are plotted below. (b) Polar niobium-oxygen N{\'e}el displacements overlaid on the BF-STEM image. Scale bar is 2 nm. The N{\'e}el niobium and oxygen along the $\mathrm{\left\langle 11\bar{2}0 \right\rangle }$ direction in green and red respectively with the solid lines referring to the averages are plotted below. (c) Curl of the niobium-oxygen displacement vector overlaid on the BF-STEM image, with the rotation vectors overlaid in white. (d) Niobium and oxygen relative Ising and N{\'e}el displacements.}
\end{figure*}

\section{\label{sec:DWDisp}Polar displacements at the bulk domain and the domain wall}

\autoref{fig:polarMap}(a) demonstrates a section of the domain wall, with the scaled Ising displacements overlaid on the corresponding BF-STEM image.  The blue regions refer to Ising displacements along $\mathrm{\left[ 000 \bar{1}\right] }$ axis, while the red regions indicate the Ising displacements along the $\mathrm{\left[ 0001 \right] }$ direction.  The $\mathrm{180^{\circ}}$ nature of the wall and the domain reversal across only one to two unit cells could be immediately ascertained, with the displacements being associated with simultaneous motion of both the niobium and the oxygen centers. Our measurements point to both oxygen and niobium atom columns displacing across the wall giving rise to a combined Ising displacement of 55 pm across the $\mathrm{180^{\circ}}$ domain wall as demonstrated in \autoref{fig:polarMap}(a).  Similar values for niobium displacements ($\mathrm{\approx 25pm}$) have been recently reported in \ce{LiNbO3} through tracking the niobium atom columns with ADF-STEM\cite{LNO_ADF}.  In addition, the wall does not maintain a sharp atomic structure showing kinks and bends along itself as shown in \autoref{fig:polarMap}(a). \autoref{fig:polarMap}(b) demonstrates that the domain wall and its proximity are also characterized by regions of N{\'e}el displacements, with parts of the wall featuring higher N{\'e}el intensities compared to the neighboring domain. In contrast to the Ising displacements, which are driven by the cooperative motion of oxygen and niobium atoms across the domain wall, the N{\'e}el displacements are however primarily driven by the niobium atoms reaching a maxima in absolute magnitude at the wall. We observe also that while the absolute magnitude of N{\'e}el displacements increase at the  domain wall, \autoref{fig:polarMap}(b) shows non-zero N{\'e}el displacements even inside the domain.  Such non-Ising displacements have been predicted before at domain wall, though one may not expect them in a hard uniaxial ferroelectric\cite{LNO_phenomenological}. Additionally, we observe that the maxima of the N{\'e}el displacements are not colocated with the center of the Ising displacements - as can be observed from \autoref{fig:polarMap}(d). This offset is due to the fact that the wall is not straight as indicated in \autoref{fig:polarMap}(a-b).  While the middle of the wall shows stronger N{\'e}el components, in the top half of the wall the N{\'e}el  displacements die out due to the slight bending of the wall.

From the Ising and N{\'e}el displacements which we map in \autoref{fig:polarMap}(a) and \autoref{fig:polarMap}(b) respectively, it is obvious that contrary to the classical expectation of a pure Ising wall, non-Ising displacements do in fact occur. This is apparent as the magnitude of the curl increases at the wall (\autoref{fig:polarMap}(c)), indicating clockwise rotation of the polar Niobium-Oxygen displacement vectors. The N{\'e}el displacements however at the domain wall have a directional preference, which may be due to the higher electrostatic energy needed for head-to-head or tail-to-tail configurations arising from bidirectional N{\'e}el displacements. The electron microscope thus paints a picture of the $180^{\circ}$ \ce{LiNbO3} domain wall where the polar displacements demonstrate variation spatially, something that we observed consistently at other images of the domain wall too (\autoref{fig:DW2}, \autoref{fig:DW3}, \autoref{fig:DW4} and \autoref{fig:DW5}), and even in images of the bulk domain (\autoref{fig:Domain1} and \autoref{fig:Domain3}), indicating perturbations in the polar order parameter, and thus increased polar entropy at the wall.

\begin{figure*}
	\centering
	\includegraphics[width=\textwidth]{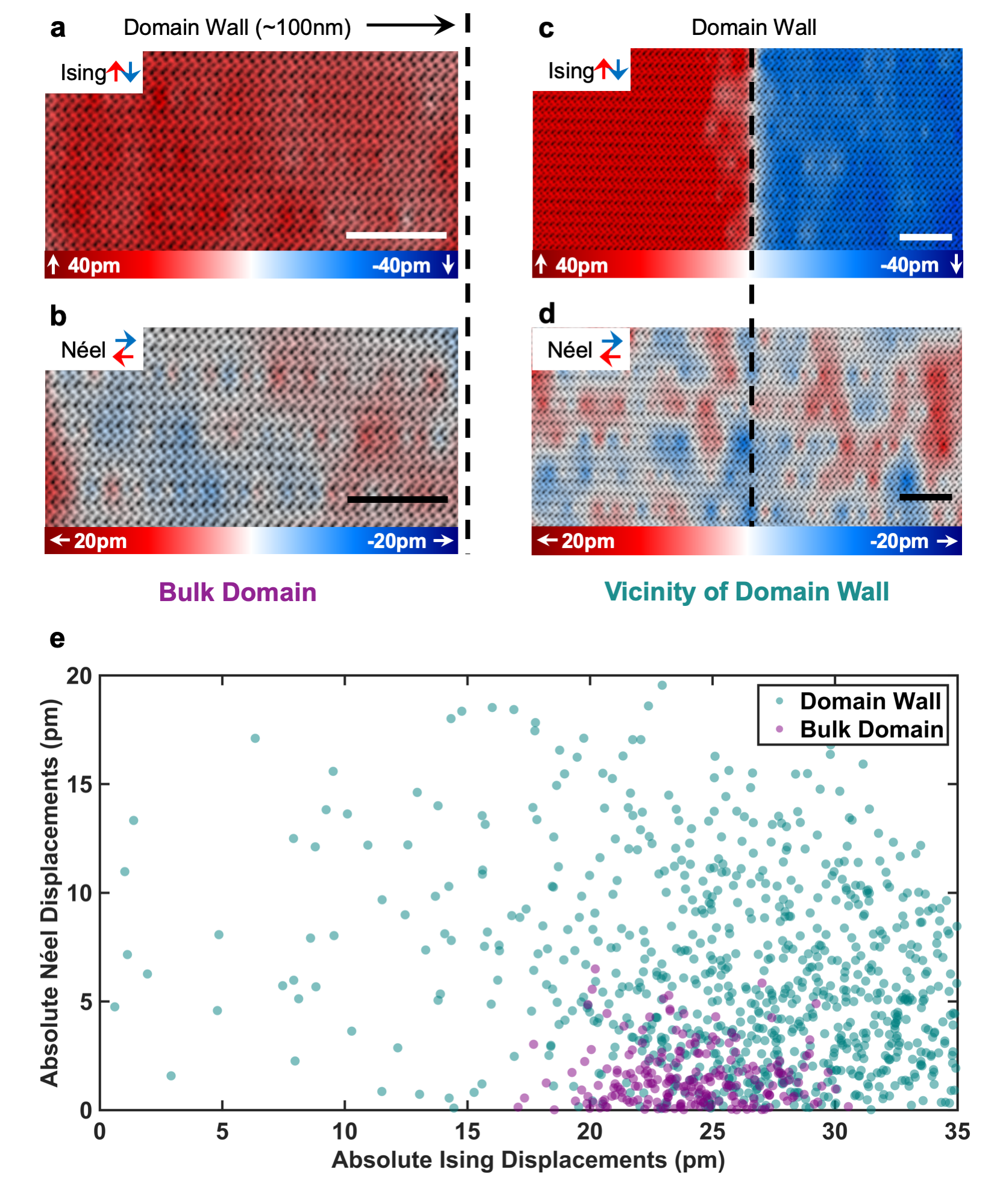}
	\caption{\label{fig:ising_neel} Comparison of polar displacements at the bulk domain versus the domain wall. (a) - (b) Ising and N{\'e}el Nb--O displacements in a region of the bulk domain, $\mathrm{\approx}$ 100nm away from the domain wall. Scale bar is 2nm. (c) - (d) Ising and N{\'e}el Nb--O displacements at the domain wall. Scale bar is 2nm. (e) Comparison of the absolute magnitudes of the Ising and N{\'e}el Nb--O displacements in the two regions imaged in \autoref{fig:ising_neel}(a)-(d) demonstrating the increased variability in displacement magnitudes at the domain wall compared to the bulk domain.}
\end{figure*}

To visualize the polar behavior away from the domain wall, we also imaged a section of the bulk domain, approximately 100nm away from the domain wall in \autoref{fig:ising_neel}(a - b). As expected, the Ising displacement direction and magnitude does not change in the bulk domain, in contrast to the Ising displacements right across the domain wall (\autoref{fig:ising_neel}(c)). Additionally we observe N{\'e}el displacements both in the bulk domain ( \autoref{fig:ising_neel}(b)) and at the domain wall ( \autoref{fig:ising_neel}(d)), with visual inspection confirming lower absolute magnitudes in the bulk domain. To understand the variability of the polar niobium--oxygen displacements at the domain wall with respect to the domain, we calculated the absolute magnitude of the Ising and N{\'e}el displacements in \autoref{fig:ising_neel}(e). These measurements demonstrate the N{\'e}el displacements reaching values below 5 pm at distances over 100 nm away from the wall, but they do not completely die down. In contrast, the absolute magnitude of the N{\'e}el displacements increases in the proximity of the domain wall -- along with a significantly higher spread in displacement magnitudes of both Ising and N{\'e}el  displacements at the domain wall compared to the bulk domain. It should be noted that both the images were obtained from the same TEM sample, and with the same exact imaging conditions. While microscope mechanical vibrations, sample preparation effects and inhomogeneities in the chemical and atomic structure can locally induce random fluctuations, this long-range decrease in the magnitude of N{\'e}el displacements (at the domain wall as opposed to the bulk domain) are probably intrinsic to \ce{LiNbO3} itself, originating from polar instabilities at the domain wall.

\section{\label{sec:Theory}First principles calculations of displacement energetics}

To understand the origin of the observed polar instabilities, we performed DFT calculations of phonons in the high symmetry phase of \ce{LiNbO3}. In agreement with previous calculations we observed three unstable modes at the $\Gamma$ point - $\mathrm{A_{2u}}$ and $\mathrm{E_u}$ polar modes with polarization parallel and perpendicular to the $\mathrm{\left[ 0001 \right] }$ direction respectively and the $\mathrm{A_{2g}}$ Raman mode (see \autoref{fig:PhononModes} and \autoref{tab:PhononModes})\cite{LNO_firstPrinciples}. The polar $\mathrm{A_{2u}}$ mode has a significant overlap with the vector representing the atomic displacements during the phase transition and therefore describes the displacement pattern responsible for the Ising macroscopic polarization of the ground state ferroelectric phase of \ce{LiNbO3}. 
 
\begin{figure*}
\centering
\includegraphics[width=\textwidth]{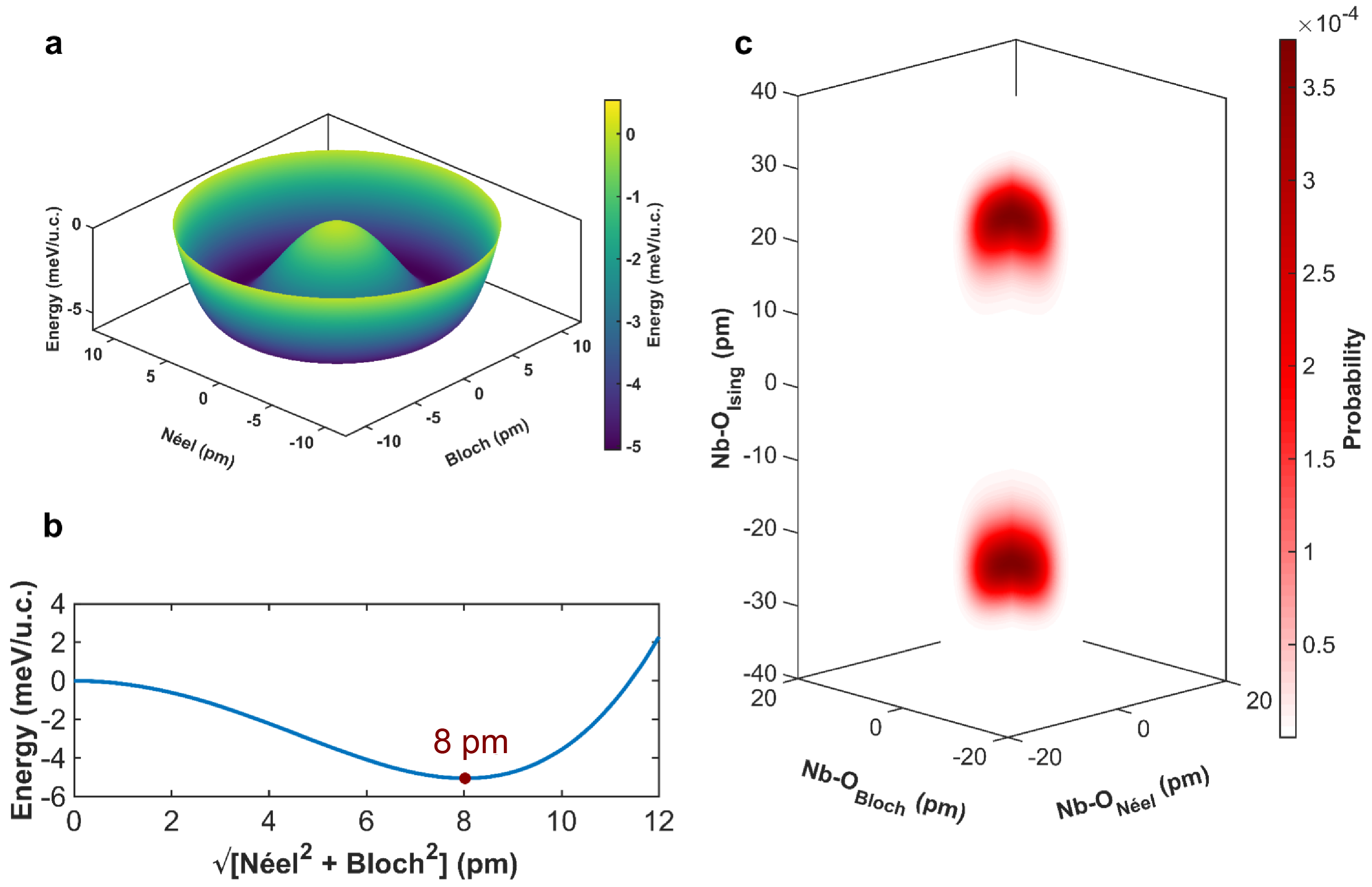}
\caption{\label{fig:theoryCalc} First Principles Calculations of displacement configurations. (a) Goldstone sombrero potential of the relative energy of \ce{LiNbO3} with polar N{\'e}el and Bloch displacements associated with the $\mathrm{E_u}$ unstable mode. (b) Energy change as a function of the combined N{\'e}el and Bloch displacement magnitude, with energy minima at 8pm. (c) Probability of displacements as a function of the Ising, N{\'e}el and Bloch displacements from mean field effective Hamiltonian.}
\end{figure*}

Moreover, we observe that the polar displacements along the N{\'e}el/Bloch direction (associated with the doubly degenerate $\mathrm{E_u}$ mode), are unstable and the system can thus lower its energy with polar displacements perpendicular to the $\mathrm{ \left[ 0001 \right] }$ Ising direction. This can explain our experimental results, i.e. the presence of N{\'e}el and Bloch polarization directions at the domain wall where the Ising polarization amplitude is strongly reduced along the Ising direction. Besides that, within a bulk ferroelectric domain, the $\mathrm{E_u}$ mode instability is suppressed by the $\mathrm{A_{2u}}$ mode condensation and the associated strain relaxation, however the energy landscape is still sufficiently shallow to allow deviations of local dipole directions from the Ising $\mathrm{ \left\langle 0001 \right\rangle }$ axis. Thus while the $\mathrm{A_{2u}}$ mode is the dominant mode driving ferroelectricity, the instability from the $\mathrm{E_u}$ modes makes it energetically favorable for the non-Ising fluctuations to arise from the ideal \ce{LiNbO3} polar configuration, thus increasing disorder in the system. 

The resultant energy landscape related to the displacements of atoms strictly perpendicular to the $\mathrm{ \left\langle 0001 \right\rangle }$ Ising axis ($\mathrm{E_u}$ mode) complies with SU(1) unitary group rotation symmetry resulting in the famous Goldstone sombrero potential shape with zero Ising component (\autoref{fig:theoryCalc}(a)).  The suppression of the Ising displacements in the $\mathrm{ \left\langle 0001 \right\rangle }$ direction thus leads to a spontaneous symmetry breakdown giving rise to the perpendicular N{\'e}el and Bloch components, with the radial magnitude of the perpendicular components reaching an energy minima at 8pm displacement (\autoref{fig:theoryCalc}(b)). We also note that the experimentally observed N{\'e}el magnitudes of approximately 10pm at the domain wall (\autoref{fig:polarMap}(d)) are close to the theoretically predicted displacement magnitude at the energy minima. Note however that these experiments cannot quantify the predicted Bloch displacements, because transmission electron microscopy probes a two-dimensional projection of columns of atoms, and Bloch displacements would be parallel to the atomic columns.  Thus it was not possible to determine whether the magnitude of the non-Ising polar components stayed constant (massless Goldstone modes) or varied across the domain boundary (massive Higgs modes)\cite{Goldstone,Higgs}. These calculations demonstrate that even in a mono-domain region, the shallow $\mathrm{E_u}$ mode permits fluctuations in the non-Ising polar components. This is shown in \autoref{fig:theoryCalc}(c) where rather than the displacements being clustered at the canonical Ising value, there is a spread in displacement magnitudes in both the N{\'e}el and the Bloch directions. Thus, our theoretical calculations demonstrate that polar disorder is intrinsic to \ce{LiNbO3} and is not just confined to the domain wall proximity.
 
\begin{figure*}
	\centering
	\includegraphics[width=\textwidth]{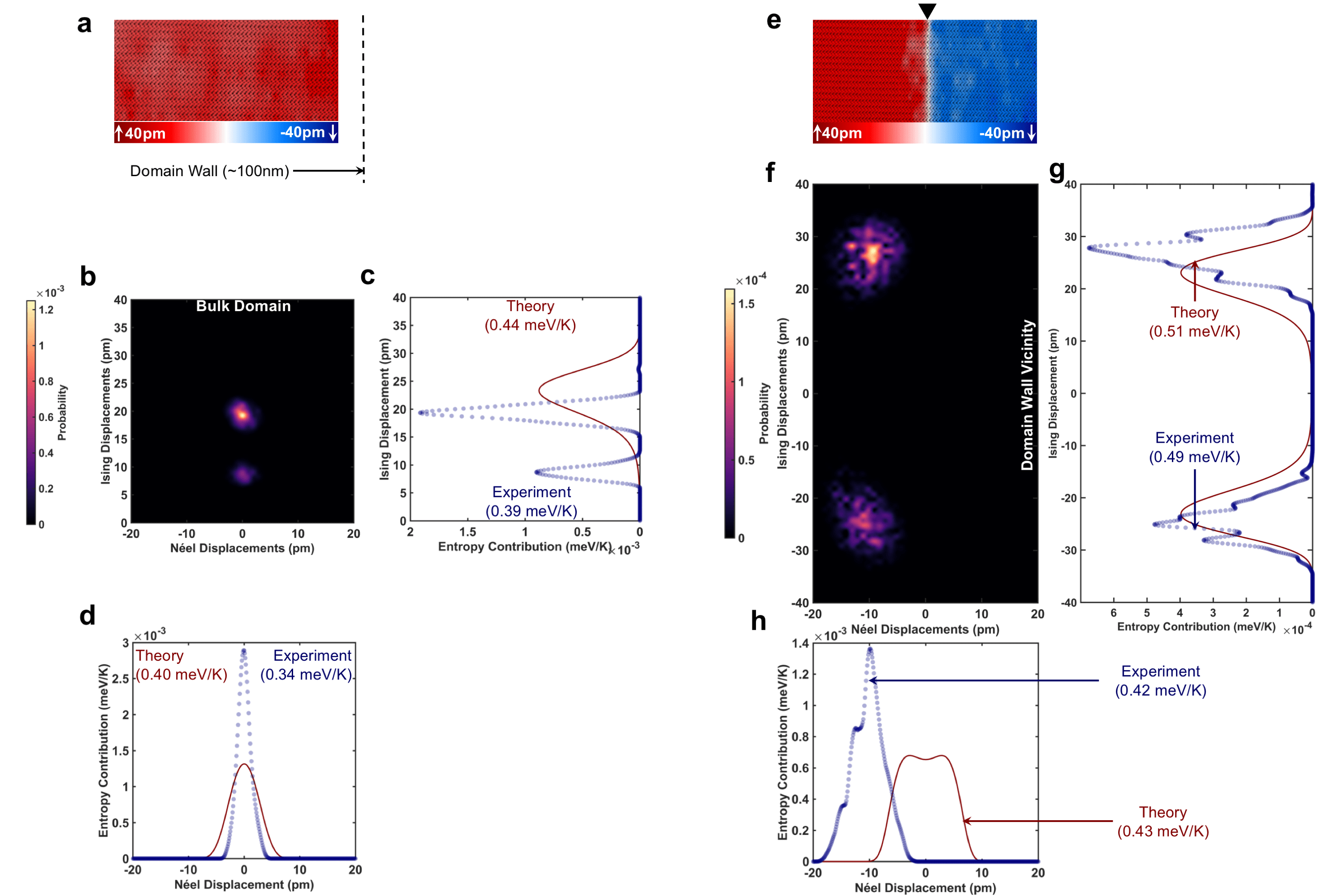}
	\caption{\label{fig:probability} Measured deconvolved probability and polar entropy in the bulk domain and domain wall proximity in \ce{LiNbO3}. (a) Representative STEM image of a bulk domain region approximately 100nm to the left of the domain wall from where the probability distribution and entropy was measured with the Ising displacements overlaid demonstrating a monodomain region. Scale bar is 2nm. (b) Richardson-Lucy deconvolved probability distribution of Ising and N{\'e}el displacement magnitudes in the bulk domain. (c) - (d) Theoretically calculated (brown) and experimentally measured (blue) entropy contribution as a function of Ising and N{\'e}el displacement orientations in the bulk domain, with the summed contribution in inset. (e) Representative STEM image of a region in the proximity of the domain wall with Ising displacements overlaid, with the black triangle marker at the top showing the domain wall location. Scale bar is 2nm. (f) Richardson-Lucy deconvolved probability distribution of Ising and N{\'e}el displacement magnitudes in the proximity of the domain wall. (g) - (h) Theoretically calculated (brown) and experimentally measured (blue) entropy contribution as a function of Ising and N{\'e}el displacement orientations in the proximity of the domain wall, with the summed contribution in inset. The entropy deconvolution and calculation process is detailed in \autoref{sec:EntropyQuantification} of the appendix.}
\end{figure*}

\section{\label{sec:DispProb}Probability distribution of polar displacements}

The standard accepted method for quantifying disorder is entropy. This can be succinctly expressed through the famous Gibbs-Boltzmann's formulation\cite{entropy_formula}:
\begin{equation}\label{eq:entropy}
S = -\sum_{N}k_b\rho \log \left( \rho \right)
\end{equation}
where $S$ is entropy, $N$ is the number of states, $k_b$ is the Boltzmann's constant, and $\rho$ is the probability of a state. Thus a single-state system has zero entropy, while entropy increases with increasing disorder, or increasing number of states. In this work, we measure polar entropy through the quantification of the probabilities of the observed polar displacements $\left( \rho \right)$ , where each possible displacement configuration is a single state. It can thus be deduced also that a monodomain system with a constant value of polar displacement has zero polar entropy.

Experimental quantification of the probabilities $\left( \rho \right)$ of polar displacements along the Ising and N{\'e}el displacement orientations in the bulk domain (representative image shown in \autoref{fig:probability}(a)) is shown in \autoref{fig:probability}(b). The measured displacements do not correspond to one single Ising value and are associated with a spread in Ising and N{\'e}el magnitudes with the most probable displacement configuration being 20 pm of Ising displacements and below 5pm of N{\'e}el displacements. The Ising displacements demonstrate a bimodal behavior originating from the fluctuations in the $\mathrm{\left\langle 0001 \right\rangle}$ intensities that were observed experimentally in the bulk domain. The origin of these fluctuations may be a consequence of local disorder, non-stoichiometry, or vacancy agglomeration in local regions. Further research is required to explain the origin of this Ising bimodal behavior within the bulk domain. Additionally, fluctuations in the electron beam may overestimate the disorder present in the system.
	
To measure the effect of the electron beam on the measured entropy, a series of Gaussian probability distributions were deconvolved from the experimentally measured displacements, and the resulting probability distributions that resulted in the lowest entropy value was chosen. Considering the deconvolution of the Gaussian probability distributions, we have calculated a standard deviation ($\sigma$) of 8.7 pm at the bulk domain and 11.7 pm at the domain wall for the microscope instability (detailed in \autoref{sec:EntropyQuantification}). It should be noted that this $\sigma$ is similar in magnitude to error distributions reported in previous STEM measurements of oxide displacements\cite{GregImaging}. The deconvolution procedure thus removes any global fluctuations in the data -- which arise notably from microscope instabilities. However, surface damage is expected to create random displacements with no long-range order as is observed in the experimental analysis.  In this experiment the sample damage was minimized by choosing a low final milling energy of 500V (see \autoref{sec:SamplePrep}).

The resulting probability distribution after the deconvolution process is used to calculate the polar entropy contribution as a function of the possible Ising and N{\'e}el states. \autoref{fig:probability}(c) demonstrates the theoretical and experimental contributions to entropy as a function of Ising displacement configurations, with the total Ising entropy being the sum of the contributions of the individual Ising displacements. Both first principles calculations and experiments demonstrate that a non-zero polar entropy originating from a spread in displacement configurations to be present even in the bulk domain. This picture is repeated even when measuring the entropy arising as a consequence of a spread of N{\'e}el displacement probabilities, as shown in \autoref{fig:probability}(d). The integrated entropy associated with the N{\'e}el component is measured as 0.31meV/K from experiment, while theory predicts an intrinsic value of 0.4meV/k.  The integrated Ising entropy (calculated by integrating the curves in panel \autoref{fig:probability}(c)) are 0.4meV/K (experiment) and 0.44 meV/K (theory). Thus both the measured and theoretically predicted Ising and  N{\'e}el contributions to the polar entropy are within 10\% of each other - indicating the intrinsic nature of these fluctuations.

This picture changes in the proximity of the domain wall (defined here as $\mathrm{\approx \pm 10nm}$ across the wall -- with the STEM image of the representative section shown in \autoref{fig:probability}(e)), whose probability distribution is plotted in \autoref{fig:probability}(f). As expected, we observe a bimodal distribution of the probable polar states in the proximity of the wall owing to two domains being imaged rather than one, with a significantly more diffuse probability distribution as compared to the probabilities measured in \autoref{fig:probability}(b). Both the integrated experimental and the theoretically predicted entropy contributions shown in \autoref{fig:probability}(g) and \autoref{fig:probability}(h) increase in the proximity of the domain wall when compared to the bulk domain. The experimentally measured polar entropy in the proximity of the wall is approximately 28\% higher than the bulk domain entropy far away from the wall. In fact, since electron microscopy measurements project a three dimensional object into a 2 dimensions, our measurements underestimate the entropy due to the absence of Bloch displacements in the calculations. This can be understood by the fact, that entropy necessarily refers to random displacements -- thus even along the Ising and N{\'e}el directions we are measuring a column averaged displacement - not the disorder of individual unit cells. This explains to a certain extent why the experimental entropy measurements are lower than their theoretically predicted values.

\section{\label{sec:Conclusions}Conclusions}

Our results presented here are the first known experimental quantification of configurational entropy of polar displacements from atomic resolution position metrology. Theoretically predicted and experimentally measured entropy reveals a classical single crystal Ising ferroelectric, hiding considerable local intrinsic disorder that is present even in the bulk domain.  This is despite \ce{LiNbO3} having only a single symmetry allowed net polarization direction, large coercive fields for domain reversal, and a high Curie temperature indicating its stability at room temperature\cite{LNO_phenomenological,LNO_curie,LNO_coercive1,LNO_coercive2}. We show that this disorder is intrinsic to ferroelectrics and can exist even in the absence of any extrinsic factors. While previous theoretical studies have demonstrated the effect of entropy in controlling polar behavior halide perovskites, here we demonstrate experimentally that entropy is considerably more prevalent.\cite{ferro_entropy_1}. Polar disorder is a highly sought after component for functional systems like piezoelectrics and electrocalorics, and our study reveals it to be present even in systems thought to be more uniform like \ce{LiNbO3}\cite{ferro_entropy_2,ferro_entropy_3,piezo1,piezo3}. The electron microscopy based metrology techniques developed here thus allow for similar studies to be performed in other systems, even beyond ferroelectrics -- allowing the electron microscope to be used not only as an imaging system, but also for atomic resolution thermodynamic quantification.

\section*{Author Contributions}

D.M., V.G. and N.A. designed the project. D.M. prepared the electron microscopy samples, and acquired the transmission electron microscopy data, assisted by K.W. D.M. developed the MATLAB subroutines for analysis. D.M., assisted by L.M., and advised by N.A. and V.G. analyzed the electron microscopy data. S.P., assisted by E.B. performed the first principles calculations. D.M., advised by N.A. and V.G. wrote the manuscript. All authors discussed the results and commented on the manuscript. 

\begin{acknowledgments}

D.M., L.M., V.G. and N.A. would like to acknowledge support from the Penn State Center for Nanoscale Sciences, an NSF MRSEC, funded under the grant number DMR-1420620. S. P. and E.B. acknowledge the supported provided by the University of Li\`{e}ge and the EU in the context of the FP7-PEOPLE-COFUND-BeIPD project, the ARC project AIMED, and the DARPA Grant No. HR0011727183-D18AP00010 (TEE Program) and the C\'{e}ci facilities funded by F.R.S-FNRS (Grant No. 2.5020.1) and Tier-1 supercomputer of the F\'{e}d\'{e}ration Wallonie-Bruxelles funded by the Walloon Region (Grant No. 1117545). D.M., V.G. and N.A. would like to acknowledge the Penn State Materials Characterization Laboratory for use of their sample preparation and electron microscopy facilities. D.M. would like to acknowledge Dr. Haiying Wang of Penn State Materials Characterization Laboratory for help with sample preparation.

\end{acknowledgments}

\appendix

\section{\label{sec:SamplePrep}Electron Microscopy of \ce{LiNbO3}}

\begin{figure*}
	\centering
	\includegraphics[width=\textwidth]{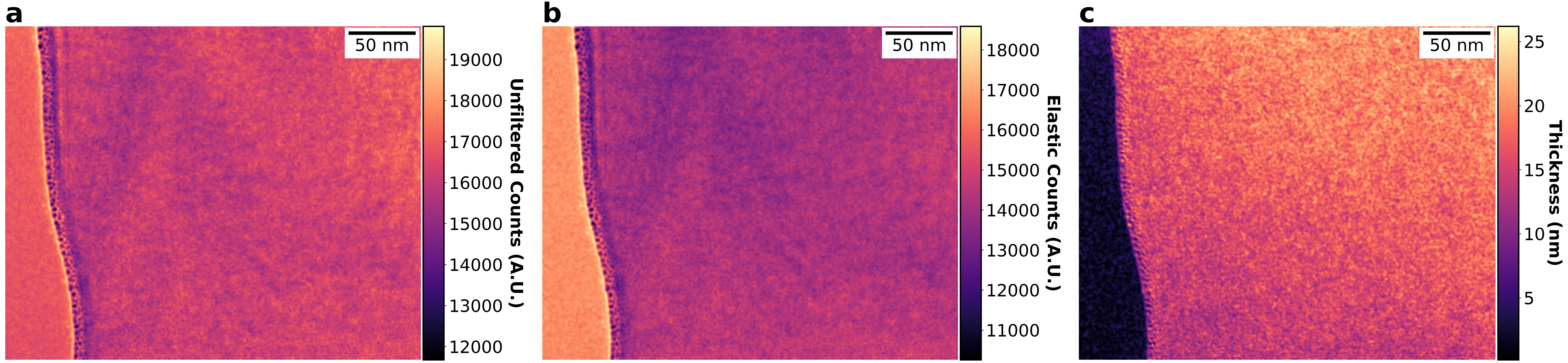}
	\caption{\label{fig:thickness} Quantification of sample thickness through EELS. (a) Low mag unfiltered (zero-loss and core-loss) EFTEM image. (b) Simultaneously acquired elastic scattering (zero-loss) EFTEM image. (c) Quantified thickness from the log of the ratio of the total inelastic and elastic (\autoref{fig:thickness}(a)) and elastic(\autoref{fig:thickness}(b)) scattering contributions, demonstrating an average thickness of approximately 20-25nm throughout the sample\cite{eels_thickness}.}
\end{figure*}

For this study, we used commercially available periodically poled single crystal congruent \ce{LiNbO3} crystals with $\mathrm{6.7\mu m}$ domain repetition from Deltronic Industries. The electron transparent samples were prepared by focused ion beam (FIB) using a FEI Helios G2 system with a 30keV gallium ion beam used for sample lift-out with the domain walls lying edge on. Final polishing was performed with 0.5kV ion beams till the sample became electron transparent at an accelerating of 2kV to ensure that the sample was thin enough for imaging oxygen atoms\cite{FIBsamplePrep}. FIB was chosen for it's advantage in site specific sample preparation. The extent of amorphous surface damage is proportional to the ion accelerating voltage - at low energies such as 2kV, the amorphous layer thicknesses are approximately 0.5--2nm thick\cite{low_FIB,FIB_damage}. A recent work has demonstrated that low voltage ion milling at 0.4kV completely eliminated amorphous surface layers\cite{FIB_amorphous}. In fact FIB has been recently used for preparing battery electrolyte TEM samples too, with no apparent damage to the sample\cite{zachman}. The prepared samples were found to have a sample thickness ranging between 20-25nm, as estimated with EELS inelastic mean free path measurements (\autoref{fig:thickness})\cite{eels_thickness}.

Following the preparation of electron transparent samples, we first imaged the \ce{LiNbO3} foil with conventional TEM (CTEM) mode at a slight defocus to locate and identify the domain walls from their diffraction contrast. Following the identification of the domain walls, we subsequently used STEM imaging in a spherical aberration corrected FEI $\mathrm{Titan^3}$ transmission electron microscope, corrected for upto third order spherical aberrations. Annular dark Field Scanning TEM (ADF-STEM) imaging was performed using Fischione detectors at a camera length of 145mm with an inner collection semi-angle of 32mrad, and an outer collection semi-angle of 188 mrad. Bright Field Scanning TEM (BF-STEM) images were simultaneously collected with Gatan detectors with an outer collection semi-angle of 15mrad. Simultaneous BF-STEM and ADF-STEM imaging was performed with fast scan directions oriented at $\mathrm{-5^{\circ}}$ and $\mathrm{85^{\circ}}$ with respect to the $\mathrm{\left\langle 11\bar{2}0 \right\rangle }$ crystallographic axis. The two image sets were combined and subsequently corrected post acquisition for scan drift using a previously developed procedure\cite{DriftCorrection}. After acquiring the atomic resolution BF-STEM images, we used MATLAB scripts to refine the positions for sub-pixel precision displacement metrology. To perform the refinement, we started by locating the highest intensity spots as a first pass to estimate atom positions in ADF-STEM and inverted contrast BF-STEM images. We performed subsequent refinement by fitting a multi-peak two-dimensional Gaussian to the observed atom intensity distribution to get the atom positions with a precision approaching $\approx 2$pm\cite{GregImaging}. 

\begin{figure}
	\centering
	\includegraphics[width=\columnwidth]{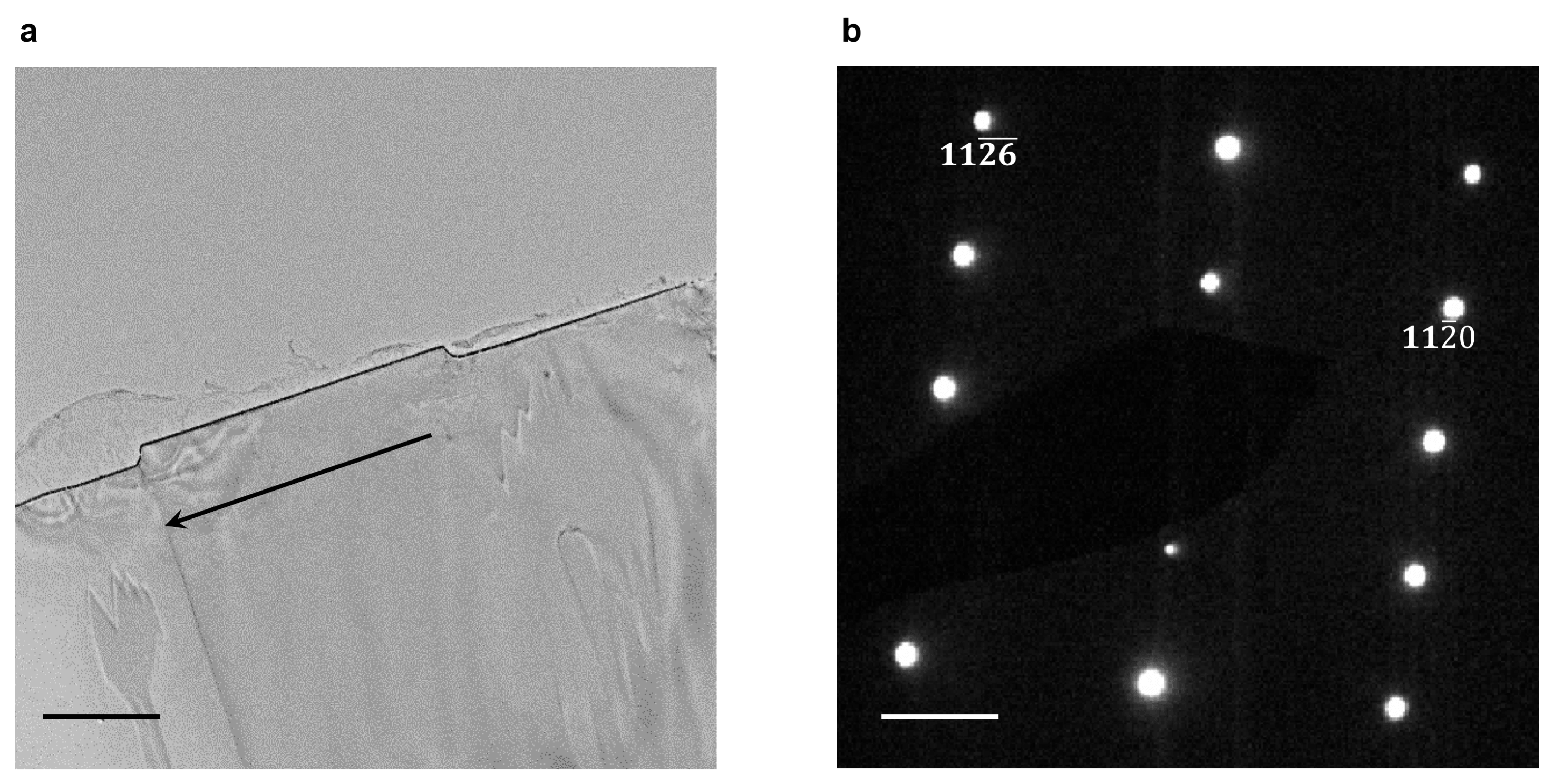}
	\caption{\label{fig:CTEM} Low magnification TEM image and electron diffraction pattern. (a) Low magnification CTEM with the domain wall (marked by the arrow) visible due to diffraction contrast at the wall. (b)  Diffraction pattern from the image in \autoref{fig:CTEM}(a) confirming the $\mathrm{\left[ 1\bar{1}00\right] }$ zone axis.}
\end{figure}

\section{\label{sec:EntropyQuantification}Quantification of Polar Entropy}
Entropy measurements are performed using STEM data acquired at both the domain wall and the bulk domain. Since the average pixel size for the experimental setup is approximately 10pm, and the approximate image size is approximately $\mathrm{2000 \times 2000}$ pixels, a representative image can visualize a $\mathrm{400nm^2}$ area. Thus images captured at the domain wall with the domain wall centered in the image field of view, still have approximately 10nm of the domain on either side. Entropy calculations are performed on one full image, and thus domain wall entropy measurements also include contributions from approximately 10nm of the domain on either side of the boundary. On average, the BF-STEM images from our experimental conditions correspond to approximately 800 unit cells. The polar displacements at each of the individual locations were subsequently sorted into 0.1pm bins, from -50pm (minimum) to 50pm (maximum) of displacement magnitude - both for Ising and N{\'e}el displacements with a total of $\mathrm{1001 \times 1001}$ possible displacement configurations. Thus, if a certain unit cell corresponds to an Ising displacement of 25.386pm, and a N{\'e}el displacement of -12.456pm, it will be assigned to the bin corresponding to the displacements of 25.3 to 25.4 Ising displacements, and -12.5 to -12.4pm of N{\'e}el displacements with one unit cell corresponding to one displacement observation. Following assignment of all the observed displacements for one full image into their respective bins, the total number of observations for each bin is divided by the total observations made for the entire image. This is the probability $\left( \rho \right)$ of observing a displacement corresponding to that bin position.

\begin{figure*}
	\centering
	\includegraphics[width=\textwidth]{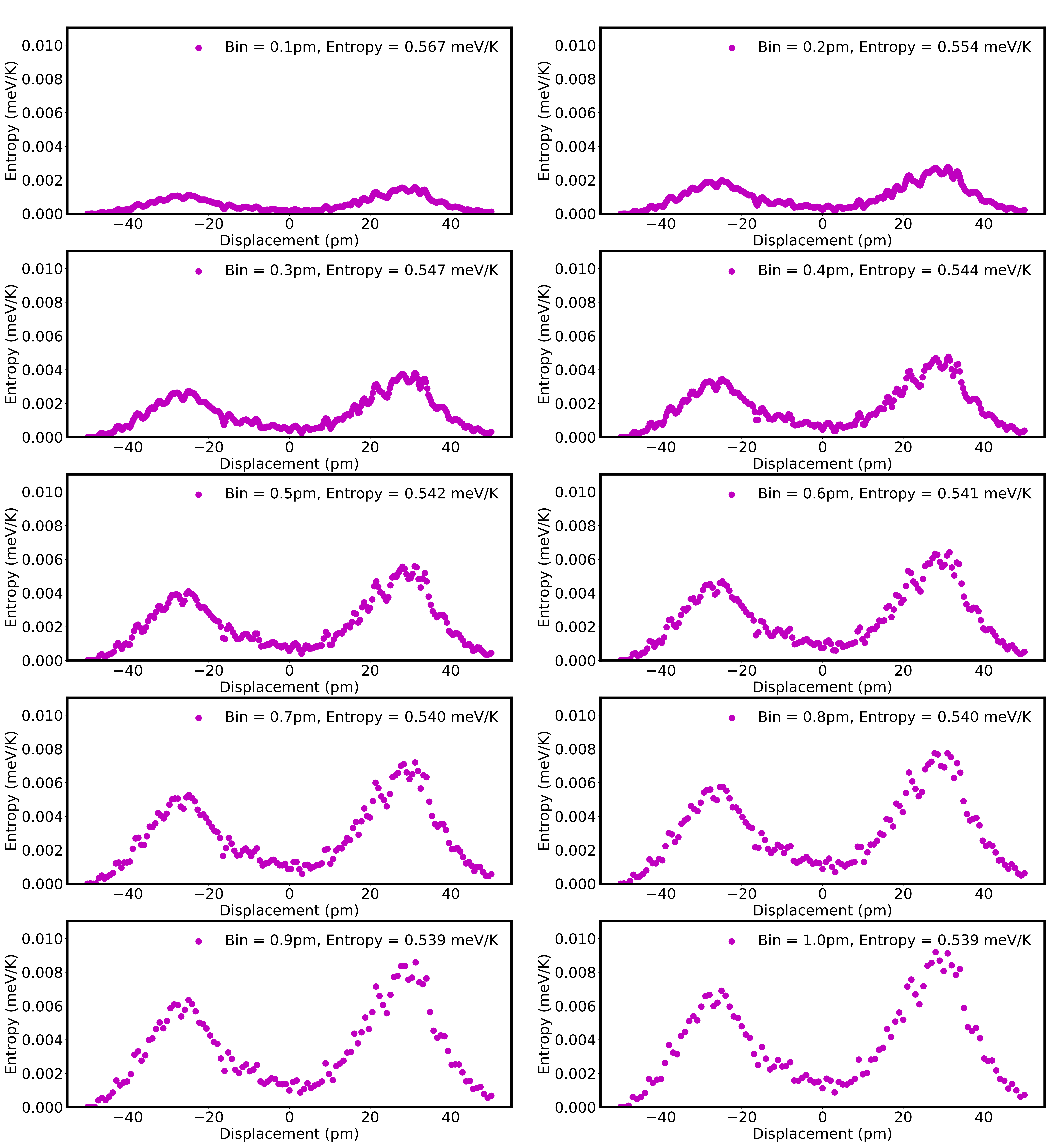}
	\caption{\label{fig:BinSize} Measured polar entropy for Ising displacements at the domain wall as a function of the displacement bin size, ranging from 0.1pm bin size to 1pm bin size.}
\end{figure*}

To quantify the effect of displacement bin size in estimating the entropy, we redid the calculations on one of the datasets -- Ising entropy calculations at the domain wall with varying bin sizes from 0.1pm to 1pm as shown in \autoref{fig:BinSize}. Choosing a 1pm bin size, rather than a 0.1pm bin size results in a reduction of the entropy from 0.567 meV/K to 0.539 meV/K, which is $\approx$ a 5\% reduction in the measured entropy. Thus, we can see that the entropy we measure is almost independent of the bin size. This can be explained by the fact that while the entropy contribution term $\left( k_b \rho\log\left( \rho \right) \right)$ in \autoref{eq:entropy} from a single displacement bin increases with increasing the bin size due to an increase of the displacement probability $\rho$ -- however this also leads to a reduction in the total number of displacement bins $\left( N \right)$, and thus the entropy which is integrated over all the possible displacement values remains fairly constant.

A shortcoming of this technique is however rooted in the fact that the measured entropy is a function of the total observed vibrational probability -- which is a combination of the \emph{intrinsic} disorder of the system itself, instabilities in the electron microscope, and induced entropy originating from the interaction between the crystal and the electron beam. While it is the intrinsic material entropy that we ideally want to measure, because of the two latter effects our measured entropy overestimates the entropy in the system. Since these measurements were performed using a single crystal of \ce{LiNbO3} where there is a remnant intrinsic entropy even in the bulk domain, there is no reference lattice to measure the microscope instabilities.

In fact even a reference lattice measurement may underestimate entropy --- for example, \ce{SrTiO3}, an ubiquitous oxide substrate is an incipient ferroelectric, with thin freestanding \ce{SrTiO3} being a ferroelectric\cite{incipientSTO}. Polar fluctuations may not be limited to \ce{LiNbO3} only, and it is highly conceivable that an entropy measurement of the substrate will also measure the intrinsic polar fluctuations of the substrate and thus overestimate microscope vibrations. 

\begin{figure*}
	\centering
	\includegraphics[width=0.75\textwidth]{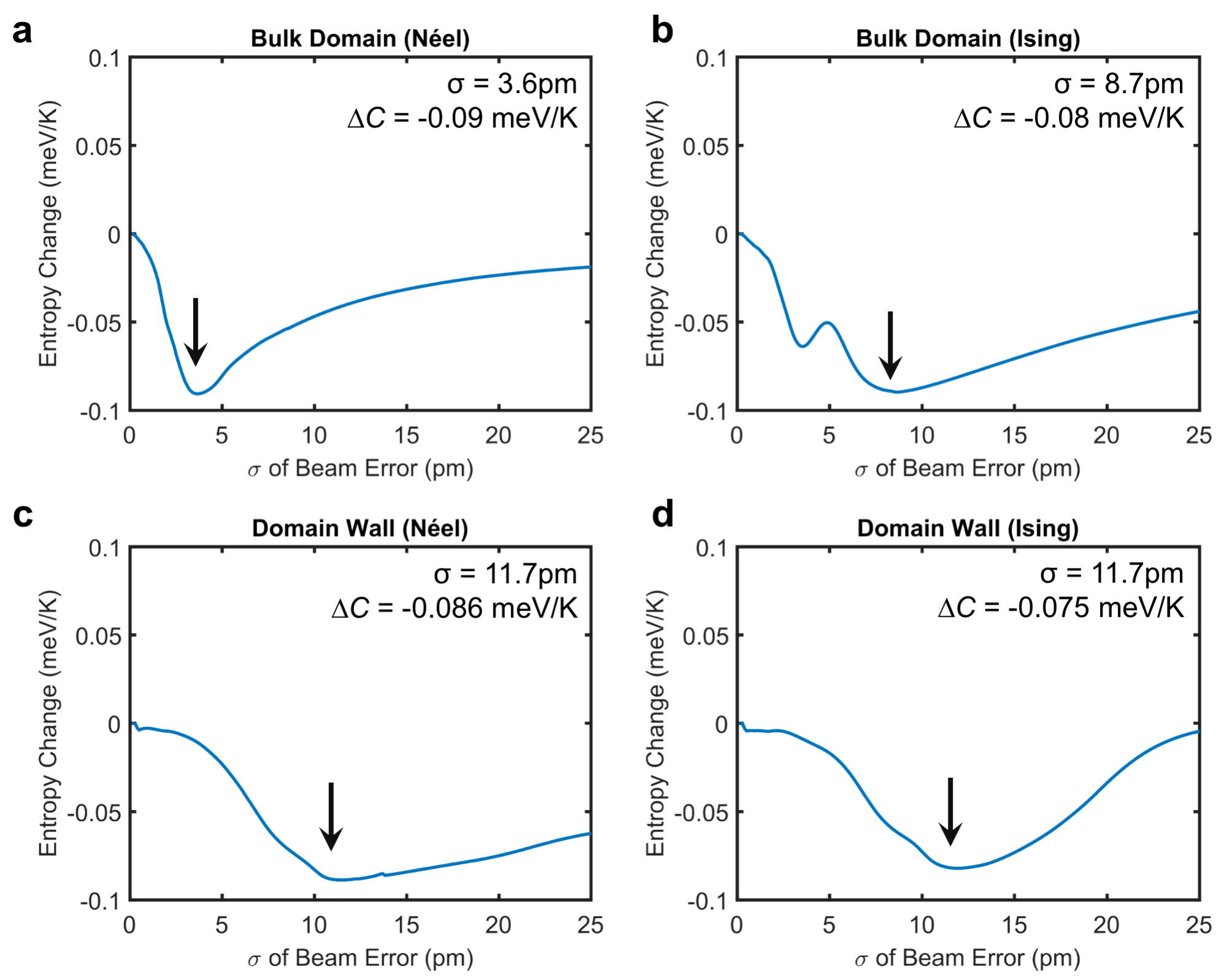}
	\caption{\label{fig:GaussianCorrection} Plotting the decrease in entropy as a function of the beam effects. (a) Change in measured entropy as a function of the measured $\sigma$ of the Gaussian of the beam probability function for N{\'e}el displacements inside the bulk domain ($\approx$ 100nm away from the domain wall.). The minima is at 3.6pm. (b) Change in measured entropy as a function of the measured $\sigma$ of the Gaussian of the beam probability function for Ising displacements inside the bulk domain ($\approx$ 100nm away from the domain wall.). There are two minima --- at 3.6pm and 8.7pm. (c) Change in measured entropy as a function of the measured $\sigma$ of the Gaussian of the beam probability function for N{\'e}el displacements in the proximity of the domain wall. The minima is at 11.7 pm. (d) Change in measured entropy as a function of the measured $\sigma$ of the Gaussian of the beam probability function for Ising displacements in the proximity of the domain wall. The minima is at 11.7 pm.}
\end{figure*}

\begin{figure*}
	\centering
	\includegraphics[width=\textwidth]{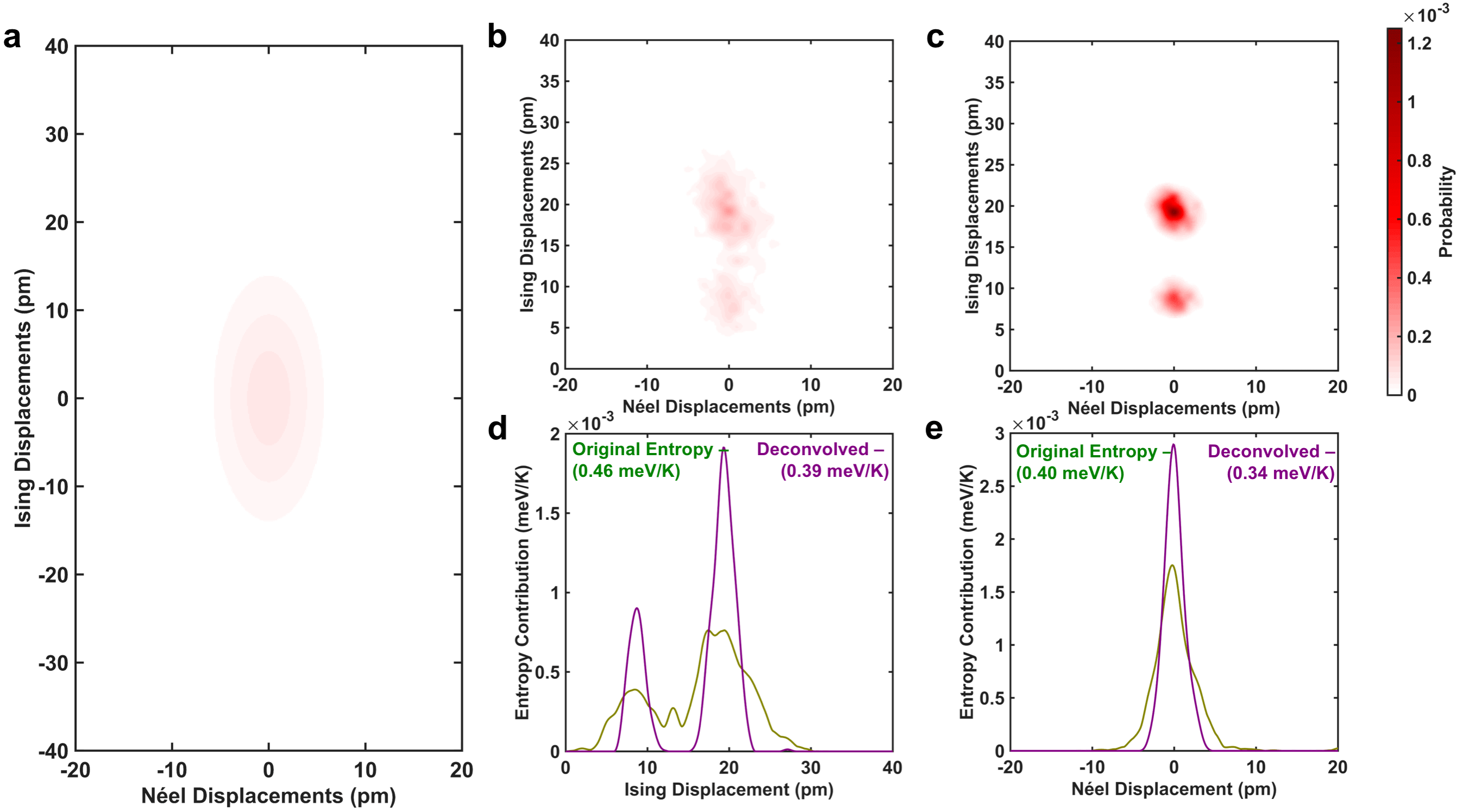}
	\caption{\label{fig:DomainDecon} Deconvoluting the microscope effects in the bulk domain. (a) Gaussian estimation of the microscope instabilities with $\mathrm{\sigma_{Neel} = 3.6pm}$ and $\mathrm{\sigma_{Ising} = 8.7pm}$. (b) Experimentally calculated probability distribution of polar displacements in the bulk domain. (c) Richardson-Lucy deconvolved probability distribution of polar displacements in the bulk domain. (d) Comparison of original and deconvolved entropy measurements as a function of Ising displacements , demonstrating an $\approx$ 13\% reduction in entropy. (e) Comparison of original and deconvolved entropy measurements as a function of N{\'e}el displacements, demonstrating an $\approx$ 23\% reduction in entropy.}
\end{figure*}  

\begin{figure*}
	\centering
	\includegraphics[width=\textwidth]{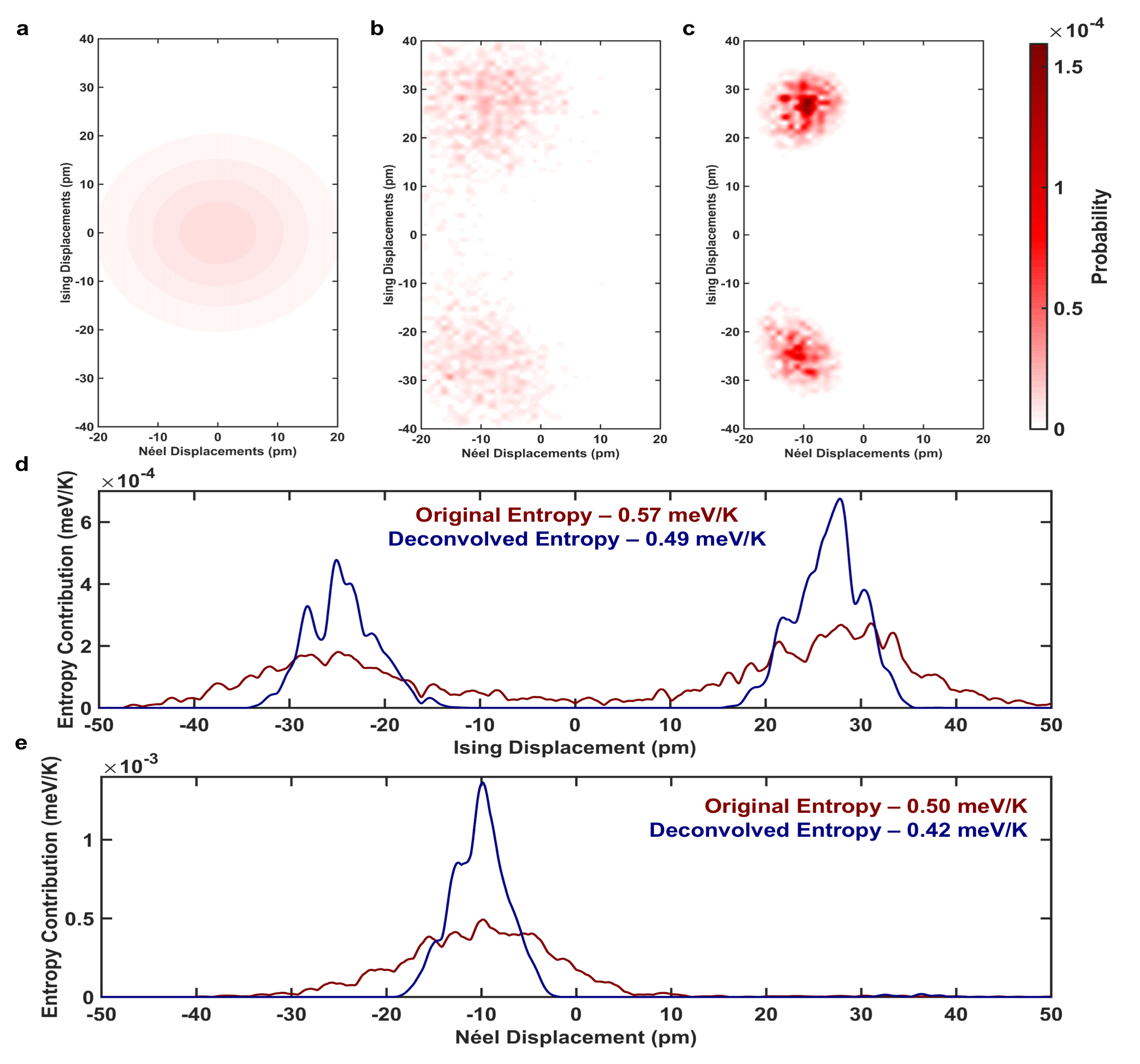}
	\caption{\label{fig:DWDecon} Deconvoluting the microscope effects in the proximity of the domain wall. (a) Gaussian estimation of the microscope instabilities with $\mathrm{\sigma_{Ising} = \sigma_{Neel} = 11.7pm}$. (b) Experimentally calculated probability distribution of polar displacements in the proximity of the domain wal(c) Richardson-Lucy deconvolved probability distribution of polar displacements in the proximity of the domain wall. (d) Comparison of original and deconvolved entropy measurements as a function of Ising displacements , demonstrating an $\approx$ 14\% reduction in entropy. (e) Comparison of original and deconvolved entropy measurements as a function of N{\'e}el displacements, demonstrating an $\approx$ 16\% reduction in entropy.}
\end{figure*}

To estimate the contribution from measurement we assume that the instabilities can be expressed as a two-dimensional Gaussian distribution with a $\mathrm{\sigma_x}$ and a $\mathrm{\sigma_y}$. The same assumption was also made for electron beam induced atom vibrations - and this is justified since the Debye-Waller parameters for both Niobium and Oxygen ( $\mathrm{u_{Nb} = 0.3924}$ and $\mathrm{u_O = 0.5}$\cite{LNO_XRD}) can be approximated as scalars rather than tensors\cite{LNO_XRD}. Since the convolution of a Gaussian kernel with another co-located Gaussian kernel is also a Gaussian, thus the total non-intrinsic microscope instability contribution $\left( \rho_{x,y} \right)$ can be reasonably approximated as a two-dimensional Gaussian function where $x$ and $y$ are the Cartesian displacement directions.

Thus, the Gaussian function can be written as:
\begin{equation} \label{eq:gaussian}
\rho_{x,y} = \frac{1}{\sqrt{2\pi}\sigma} e^{ - \frac{1}{2} \left( \left( \frac{x - \mu_x}{\sigma_x} \right)^2 + \left( \frac{y - \mu_y}{\sigma_y} \right)^2 \right) } \delta x \delta y
\end{equation}

This is a probability distribution, since:
\begin{align}
\int_{-\infty}^{\infty} \int_{-\infty}^{\infty} \rho_{x,y} & = \int_{-\infty}^{\infty} \int_{-\infty}^{\infty} \frac{1}{\sqrt{2\pi}\sigma} e^{ - \frac{1}{2} \left( \left( \frac{x}{\sigma_x} \right)^2 + \left( \frac{y}{\sigma_y} \right)^2 \right) } \delta x \delta y \\
& = 1
\end{align}

Since this is the microscope instability probability, the Gaussian is centered at $\left( \mu_x,\mu_y \right) = 0$, then \autoref{eq:gaussian} can be expressed as:
\begin{equation} \label{eq:gauss}
\rho_{x,y} = \frac{1}{\sqrt{2\pi}\sigma} e^{ - \frac{1}{2} \left( \left( \frac{x}{\sigma_x} \right)^2 + \left( \frac{y}{\sigma_y} \right)^2 \right) } \delta x \delta y
\end{equation}. 

Using the Boltzmann definition of entropy, and inputing \autoref{eq:gauss} in \autoref{eq:entropy}
\begin{equation}\label{eq:gaussian_entropy}
\Delta S_e = k_B \left[ \log{\left(  \sqrt{2\pi}\sigma_x + \sigma_y \right) } + \frac{1}{2}   \right] 
\end{equation}

where $\Delta S_e$ refers to the microscope contribution to measured entropy.

Thus, as \autoref{eq:gaussian_entropy} demonstrates, the total entropy increases monotonically with $\sigma_x$ and $\sigma_y$. To obtain the intrinsic probability, a deconvolution of the measured probability distribution function with a Gaussian PDF is thus required. It can be easily deduced that the microscope contribution to the entropy is mathematically analogous to a blurring function commonly encountered in optics. Thus, the experimentally measured probability is the material probability convolved by a point spread function (PSF) of the instrumental vibrations, with the PSF assumed to be Gaussian in this case. To deconvolve the underlying entropy, we can thus use the Richardson-Lucy deconvolution to iteratively obtain the unblurred PDF\cite{RLucy1,RLucy2}. Mathematically, thus if $\rho$ is the microscope instability probability distribution, and $\tau$ is the intrinsic fluctuation in polar displacements, it is the following entropy we are after. 
\begin{equation}
\Delta S_i = \int k_B \tau\log \tau
\end{equation}
Since, the measured probability distribution $\phi$ is a convolution, it can be written as:
\begin{align}
\phi & = \theta \circ \rho\\
& = \mathcal{F}\left( \mathcal{F}^{-1}\left( \tau \right) \times \mathcal{F}^{-1}\left( \rho \right) \right)
\end{align}
Thus for a certain value of $\sigma_x$ and $\sigma_y$, the deconvolved probability as a function of $\sigma_x$ and $\sigma_y$ can be expressed as
\begin{equation}
\tau_{\sigma_x,\sigma_y} = \mathcal{F}\left( \frac{\mathcal{F}^{-1}\left(\phi\right) \times \left(\mathcal{F}^{-1}\left(\rho_{\sigma_x,\sigma_y}\right)\right)^*}{| \mathcal{F}^{-1}\left( \rho_{\sigma_x,\sigma_y} \right) |^2} \right)
\end{equation}
where $\mathcal{C}^*$ is the complex conjugate of a function $\mathcal{C}$. 

Thus, the decrease in entropy $\Delta C$ as a result of the deconvolution :
\begin{equation}\label{eq:entropy_decrease}
\Delta C_{\sigma_x,\sigma_y} = -k_B \left( \int \phi \log \phi - \int \tau_{\sigma_x,\sigma_y} \log \tau_{\sigma_x,\sigma_y} \right)
\end{equation}

Since we do not possess a reference lattice from which a point spread function could be deduced, we measured $\Delta C_{\sigma_x,\sigma_y}$ for both Ising and N{\'e}el displacements in the bulk domain and at the domain wall. For N{\'e}el displacements in the bulk domain, the $\Delta C_{\sigma_x,\sigma_y}$ reaches a minima with a $\mathrm{\sigma_x = 3.6pm}$ as demonstrated in \autoref{fig:GaussianCorrection}(a). Qualitatively, this means that the PDF corresponding to this particular displacement can be most closely be approximated a Gaussian of  $\mathrm{\sigma_x = 3.6pm}$. Compared to the N{\'e}el displacements, when we plot $\Delta C_{\sigma_x,\sigma_y}$ for Ising displacements in the bulk domain in \autoref{fig:GaussianCorrection}(b), we encounter two minima - one identical to the minima in \autoref{fig:GaussianCorrection}(a) at 3.6pm, and the second minima at 8.7pm. The origin of this behavior could be understood by looking at the probability distribution of Ising displacements in the bulk domain (\autoref{fig:probability}(b) \& \autoref{fig:probability}(c)) which is bimodal. For the most conservative possible estimate, thus all domain probabilities were calculated after deconvolving the measured probability distribution with a Gaussian of $\sigma_x = 3.6$pm and $\sigma_y = 8.7$pm.

Extending the deconvolution to the proximity of the domain wall, we observe that the maximum decrease in entropy occurs for both N{\'e}el and Ising displacements when the measured probability distribution function is deconvolved with a Gaussian with $\mathrm{\sigma = 11.7pm}$. It is interesting to note that the $\sigma_x$ and $\sigma_y$ is larger in the proximity of the domain wall, than in the bulk domain. However, since the experimental data for both regions was acquired back to back in the same experimental session - it is highly unlikely that the microscope is quantifiably less stable at the domain wall than at domain. Rather, any Gaussian features in the probability distribution function are in fact being assigned to the point spread function, and thus this technique of measuring entropy is actually slightly conservative -- the deconvolved entropy is in reality \emph{underestimating} the intrinsic material entropy.

Visually, we can understand the effect of the deconvolution by observing the $\sigma_x = 3.6$pm and $\sigma_y = 8.7$pm Gaussian distribution in \autoref{fig:DomainDecon}(a), the original measured probability distribution in \autoref{fig:DomainDecon}(b) and the deconvolved probability in \autoref{fig:DomainDecon}(c) in the bulk domain. The deconvolved probability is significantly sharper and less spread out. This is borne out by a reduction of the Ising contribution of the entropy from 0.46meV/K to 0.40meV/K, a reduction of $\approx$ 13\%, as demonstrated in \autoref{fig:DomainDecon}(d). The N{\'e}el  contribution to the entropy, plotted in \autoref{fig:DomainDecon}(e) also declines by $\approx$ 23\% from 0.40 meV/K to 0.31meV/K.

Similarly, plotting the effects of the $\mathrm{\sigma_x,\sigma_y = 11.7pm}$ Gaussian point spread function in the proximity of the domain wall, we find a marked sharpening of the deconvolved probability distribution function in \autoref{fig:DWDecon}(c) when compared to the experimentally measured probability distribution in \autoref{fig:DWDecon}(b). This sharpening leads to a reduction of the Ising contribution to the entropy, as plotted in \autoref{fig:DWDecon}(d) by 14\% from 0.57meV/K to 0.49meV/K. The N{\'e}el  contribution to the entropy, plotted in \autoref{fig:DWDecon}(e), declines by 16\% as a result of the deconvolution from 0.50 meV/K to 0.42 meV/K. Thus, while the deconvolution decreases the total measured entropy across the board, even using the most aggressive Gaussian kernel does not result in zero entropy, showing that this polar entropy is intrinsic to the material itself.

Thus, we observe around $\approx$ 25\% reduction in the measured entropy due to the deconvolution. The entropy of displacements $\left( S \right)$ were subsequently calculated from the deconvolved  probability distributions as per \autoref{eq:entropy}.

\section{\label{sec:fpcalc}First principles calculations}

First principles calculations were done using the density functional theory approximation as implemented in the ABINIT software package (v.8.4.3)\cite{ABINIT1,DFTbook,KohnHohenberg,ABINIT2}. We chose the libxc implementation of PBEsol GGA functional to describe the exchange-correlation energy contribution, and the valence electrons were teated through norm-conserving pseudopotentials obtained through the PseudoDojo project\cite{DFTgradient,libxc,PseudoDojo,Vanderbilt,Zformula}. The planewave kinetic cut-off energy was taken to be equal to 50 Ha and the Brillouin zone was sampled using a $\mathrm{6\times 6\times 6}$ Monkhorst-Pack mesh of special $k$ points\cite{monkhorst}. To determine the structure of the paraelectric $\mathrm{R\bar{3}c}$ phase structure of \ce{LiNbO3}, we considered a primitive 10 atom unit cell and performed a relaxation of atomic positions followed by an energy optimization with respect to changes both in lattice vectors and the reduced atomic coordinates under an imposed constraint of the fixed $\mathrm{R\bar{3}c}$ space-group symmetry , with the primitive unit cell dimensions given in \autoref{tab:Paraelectric}. The high-accuracy structural relaxation was performed until the calculated force magnitudes were less than $\mathrm{10^{-8}ev}$ per \AA, and the absolute values of stress tensor components do not exceed $\mathrm{10^{-7} GPa}$. We performed density functional perturbation theory calculations (DFPT) so as to identify the unstable phonon modes (\autoref{tab:PhononModes}. To construct the minimal effective Hamiltonian model we have first computed the internal energy landscapes for all identified unstable modes. For this, we have performed DFT calculations of the total energy change upon gradually condensing the unstable modes into the structure. The resulting curves were fitted with the $\mathrm{8^{th}}$ order polynomials as given by \autoref{eq:polynomial}.

\begin{equation}\label{eq:polynomial}
E_M = \kappa_Mx^2 + \alpha_Mx^4 + \gamma_Mx^6 + \delta_Mx^8
\end{equation}
where x denotes the amplitude of the mode M. Similarly, performing calculations of energy changes induced by displacements involving not a single but two phonon modes allows to reconstruct the effective mode interactions that we take here to be of the form 
\begin{equation}
E_{int}^{M_1M_2} = gM_1M_2x^2y^2
\end{equation}

where x and y denote the amplitudes of the $M_1$ and $M_2$ modes. The interaction of local modes with strain is taken into account by fitting the dependences of elastic stresses on the mode amplitudes. Finally, the elastic energy produced by the deformations of the cell shape and volume is taken into account in the harmonic approximation. The elastic constants are computed from density functional perturbation theory. Note that in the case of the Eu mode, all the energy expansion coefficients are assumed to depend on the displacement direction in the (0001) plane, however the calculations show that such in-plane anisotropy can be safely neglected. In the described model, the short-range and long-range dipolar interactions between different modes are taken into account in the mean-field approximation -- these energetic contributions essentially lead to renormalization of the $\kappa_M$  and $f_{M_1 M_2}$ coefficients. To determine the most important low-energy atomic displacements patterns we further performed the density functional perturbation theory calculations so as to identify low frequency phonon modes for the obtained ground state.

\section{\label{sec:PolarModes}Calculation of polar modes}

\begin{table}
	\centering
	\begin{tabular}{ll}
		\toprule
		\textbf{Atom} & \textbf{Position}\\
		\midrule
		Li1 & $\mathrm{\left( 0,0,\sfrac{1}{2} \right)}$\\
		Nb1 & $\mathrm{\left( 0,0,0 \right)}$\\
		O1 & $\mathrm{\left( \sfrac{-1}{3},\sfrac{-1}{3}+x,\sfrac{7}{12} \right)}$\\
		O2 & $\mathrm{\left( \sfrac{1}{3}-x,-x,\sfrac{7}{12}\right) }$\\
		O3 & $\mathrm{\left(x,\sfrac{1}{3},\sfrac{7}{12}\right) }$\\
		\bottomrule
	\end{tabular}
	\caption{\label{tab:Paraelectric}Calculated hexagonal co-ordinates of atoms of the primitive unit cell of paraelectric \ce{LiNbO3}. The lattice parameters are a  = 518pm and c = 1364.6pm.}
\end{table}

\begin{figure*}
	\centering
	\includegraphics[width=0.75\textwidth]{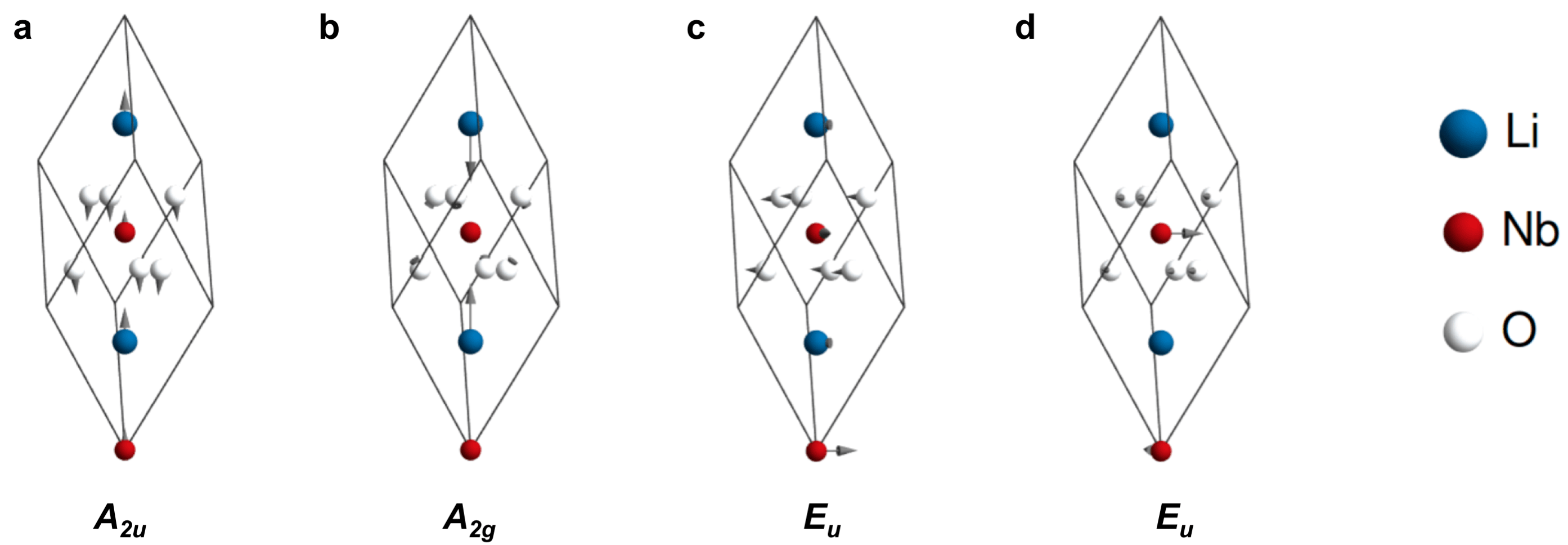}
	\caption{\label{fig:PhononModes} Eigenvectors of the phonon modes. (a) the $\mathrm{A_{2u}}$ mode. (b) the $\mathrm{A_{2g}}$ mode. (c) and (d) the two eigenvectors corresponding to the degenerate $\mathrm{E_u}$ mode aligned with x and y Cartesian axis respectively.}
\end{figure*}

The calculated polar modes for the paraelectric \ce{LiNbO3} unit cell (\autoref{tab:Paraelectric}) are shown in \autoref{tab:PhononModes}. As could be observed, there are four polar modes, with the $\mathrm{A_{2u}}$ mode driving ferroelectricity, while it is the degenerate $\mathrm{E_u}$ modes that drive the non-Ising N{\'e}el and Bloch displacements. The polar phonon mode displacements are visualized in \autoref{fig:PhononModes}, which plot the individual atom displacements corresponding to the polar modes. 

\begin{table}
	\centering
	\begin{tabular}{lcccc}
		\toprule
		\textbf{Atom} & \multicolumn{4}{c}{\textbf{Polar Phonon Modes}}\\
		\cmidrule(l){2-5}
		& $\mathbf{A_{2g}}$ & $\mathbf{A_{2u}}$ & $\mathbf{E_u}$ & $\mathbf{E_u}$ \\
		\midrule
		Li1 & $\begin{pmatrix} 0.000 \\ 0.000 \\0.407 \end{pmatrix}$ & $\begin{pmatrix} 0.000 \\ 0.000 \\ 0.683 \end{pmatrix}$ & $\begin{pmatrix} 0.000 \\ -0.069 \\ 0.000 \end{pmatrix}$ & $\begin{pmatrix} 0.069 \\ 0.000 \\ 0.000 \end{pmatrix}$ \\
		Li2 & $\begin{pmatrix} 0.000 \\ 0.000 \\ 0.407 \end{pmatrix}$ & $\begin{pmatrix} 0.000 \\ 0.000 \\ -0.683 \end{pmatrix}$	 & $\begin{pmatrix} 0.000 \\ -0.069 \\ 0.000 \end{pmatrix}$ & $\begin{pmatrix} 0.069 \\ 0.000 \\ 0.000 \end{pmatrix}$ \\
		Nb1 & $\begin{pmatrix}0.000 \\ 0.000 \\ 0.259 \end{pmatrix}$ & $\begin{pmatrix} 0.000 \\ 0.000 \\ 0.000 \end{pmatrix}$ & $\begin{pmatrix} -0.384 \\ 0.356 \\ 0.259 \end{pmatrix}$ & $\begin{pmatrix} -0.356 \\ -0.384 \\ 0.000 \end{pmatrix}$ \\
		Nb2 & $\begin{pmatrix} 0.000 \\ 0.000 \\ 0.259 \end{pmatrix}$ & $\begin{pmatrix} 0.000 \\ 0.000 \\ 0.000 \end{pmatrix}$ & $\begin{pmatrix} 0.384 \\ 0.356 \\ 0.000 \end{pmatrix}$ & $\begin{pmatrix} -0.356 \\ 0.384 \\ 0.000 \end{pmatrix}$ \\
		O1 & $\begin{pmatrix} 0.029 \\ 0.017 \\ -0.297 \end{pmatrix}$ & $\begin{pmatrix} 0.055 \\ 0.032 \\ -0.085 \end{pmatrix}$	 & $\begin{pmatrix} 0.002 \\ -0.272 \\ 0.012 \end{pmatrix}$ & $\begin{pmatrix} 0.269 \\ -0.002 \\ -0.021 \end{pmatrix}$ \\
		O2 & $\begin{pmatrix} -0.029 \\ 0.017 \\ -0.297 \end{pmatrix}$	 & $\begin{pmatrix} -0.055 \\ -0.032 \\ 0.085 \end{pmatrix}$	 & $\begin{pmatrix} -0.002 \\ -0.272 \\ -0.012 \end{pmatrix}$ & $\begin{pmatrix} 0.269 \\ 0.002 \\ -0.021 \end{pmatrix}$ \\
		O3 & $\begin{pmatrix} 0.000 \\ -0.033 \\ -0.297 \end{pmatrix}$ & $\begin{pmatrix} 0.000 \\ -0.064 \\ -0.085 \end{pmatrix}$ & $\begin{pmatrix} 0.000 \\ -0.268 \\ -0.024 \end{pmatrix}$ & $\begin{pmatrix} 0.274 \\ 0.000 \\ 0.000  \end{pmatrix}$ \\
		O4 & $\begin{pmatrix} 0.000 \\ -0.033 \\ -0.297 \end{pmatrix}$ & $\begin{pmatrix} 0.000 \\ 0.064 \\ 0.085 \end{pmatrix}$ & $\begin{pmatrix} 0.000 \\ -0.268 \\ -0.024 \end{pmatrix}$ & $\begin{pmatrix} 0.274 \\ 0.000 \\ 0.000  \end{pmatrix}$ \\
		O5 & $\begin{pmatrix} -0.029 \\ 0.017 \\ -0.297 \end{pmatrix}$ & $\begin{pmatrix} -0.055 \\ 0.032 \\ -0.085 \end{pmatrix}$ & $\begin{pmatrix} -0.002 \\ -0.272 \\ 0.012 \end{pmatrix}$ & $\begin{pmatrix} 0.269 \\ 0.002 \\ 0.021 \end{pmatrix}$ \\
		O6 & $\begin{pmatrix} 0.029 \\ 0.017 \\ -0.297 \end{pmatrix}$ & $\begin{pmatrix} -0.055 \\ -0.032 \\ 0.085 \end{pmatrix}$ & $\begin{pmatrix} 0.002 \\ -0.272 \\ 0.012 \end{pmatrix}$ & $\begin{pmatrix} 0.269 \\ -0.002 \\ -0.021  \end{pmatrix}$ \\
		\bottomrule
	\end{tabular}
	\caption{\label{tab:PhononModes} Eigenvectors in Cartesian coordinates of the identified unstable phonon modes of paraelectric \ce{LiNbO3}}
\end{table}

\section{\label{sec:DisplacementsBulk}Displacements in the Bulk Domain}

\begin{figure*}
	\centering
	\includegraphics[width=0.75\textwidth]{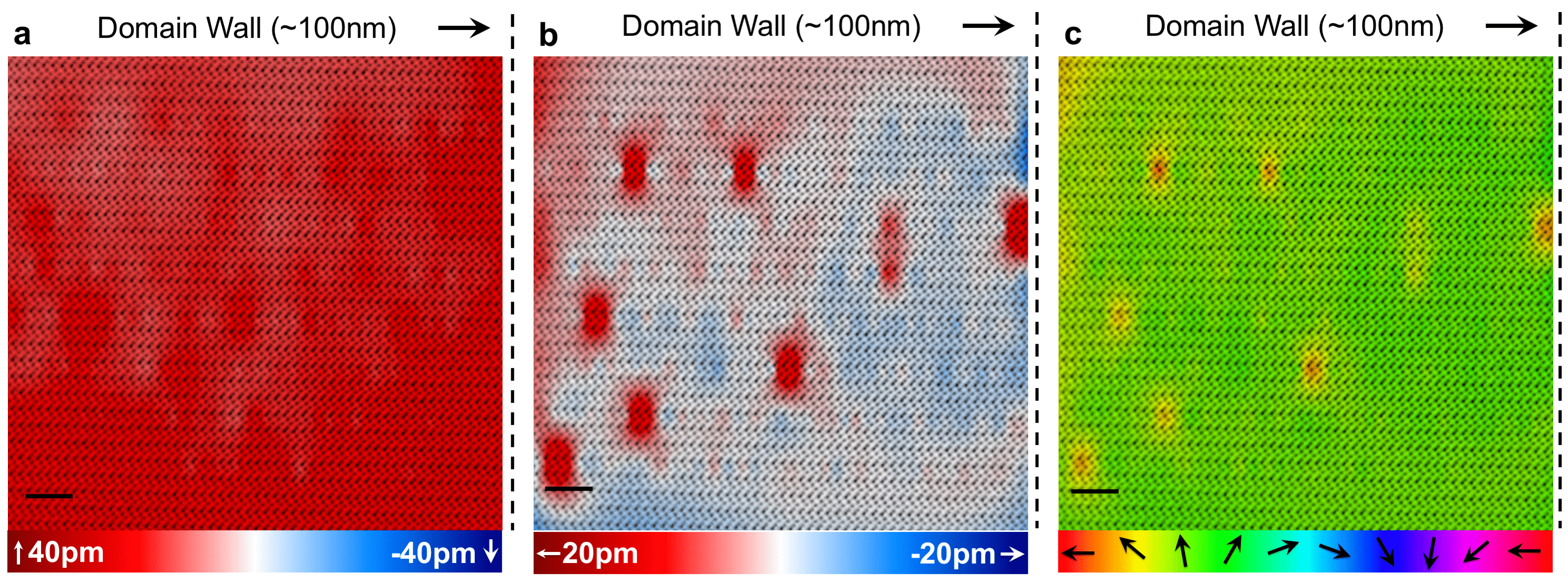}
	\caption{\label{fig:Domain1} Bulk domain HR-STEM image with the  polarization and rotation map overlaid on top at a location $\mathrm{\approx 100nm}$ away from wall. (a) Ising displacement mapped out over the bulk domain. (b) N{\'e}el displacement map showing regions of no N{\'e}el displacements, and nanoregions of high N{\'e}el displacements. (c) Rotation colormap. Scale bar in all images is 2nm.}
\end{figure*}

\begin{figure*}
	\centering
	\includegraphics[width=0.75\textwidth]{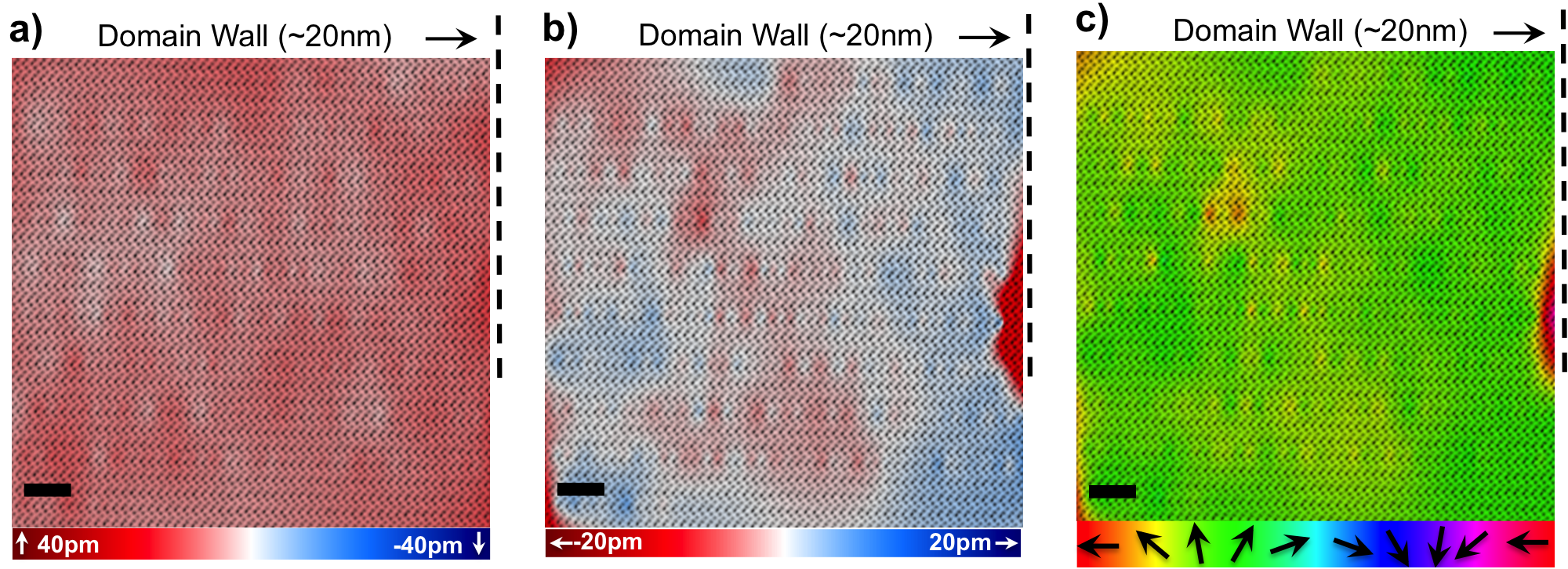}
	\caption{\label{fig:Domain3} Bulk domain $\mathrm{\approx 20nm}$ away from wall. (a) Ising displacement mapped out over the bulk domain, showing significantly lower displacements in comparison to other imaged regions. (b) N{\'e}el displacement map showing regions of lower N{\'e}el displacements. (c) Rotation color map. Scale bar in all images is 2nm}
\end{figure*}

Two different regions (\autoref{fig:Domain1} and \autoref{fig:Domain3}) are shown as different regions of the bulk domain that were imaged. While all three are mono-domain regions, it is instructive to note that the Ising displacement itself is not entirely constant even 100nm into the domain, with the displacement demonstrating magnitude variations as seen in \autoref{fig:Domain1}(a). These regions are additionally associated with regions of N{\'e}el displacements as can be observed in \autoref{fig:Domain1}(b). These N{\'e}el displacements are ultimately visible in the rotation map (see \autoref{fig:Domain1}(c)), demonstrating polar non-Ising components arising even in bulk domain regions approximately 100nm away from the domain wall. This variation in polar components is ultimately reflected in increased entropy.

\autoref{fig:Domain3} demonstrates a section of the bulk domain, approximately 20nm away from the domain wall. As could be observed in this section, the total Ising displacements are significantly smaller than expected, with a corresponding decrease in N{\'e}el displacements, demonstrating regions of decreased polarity embedded in the domain near the domain wall.

\section{\label{sec:OtherDWs}Displacements in the the proximity of the Domain Wall}

\begin{figure}
	\centering
	\includegraphics[width=\columnwidth]{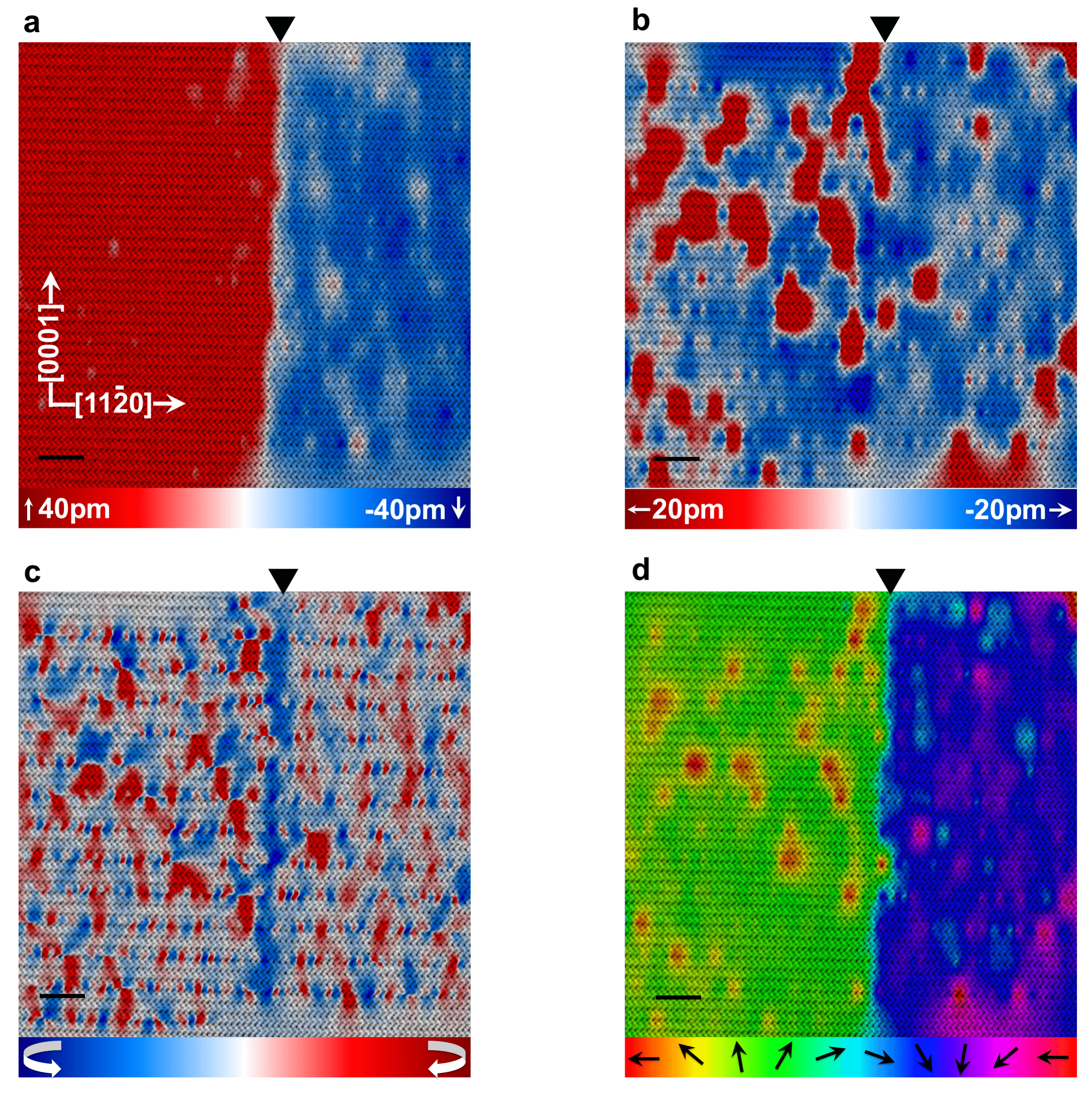}
	\caption{\label{fig:DW2} Domain wall in Region 2. (a) Ising displacements at region 2 of the domain wall with non-equivalent polarization on either sides. (b) N{\'e}el displacements demonstrating the presence of strong alternating N{\'e}el components. (c) Curl of the polar niobium-oxygen displacement map with slight decrease at the domain wall. (d) Rotation map of the polar niobium-oxygen displacements. Scale bar in all images is 2nm.}
\end{figure}

\begin{figure}
	\centering
	\includegraphics[width=\columnwidth]{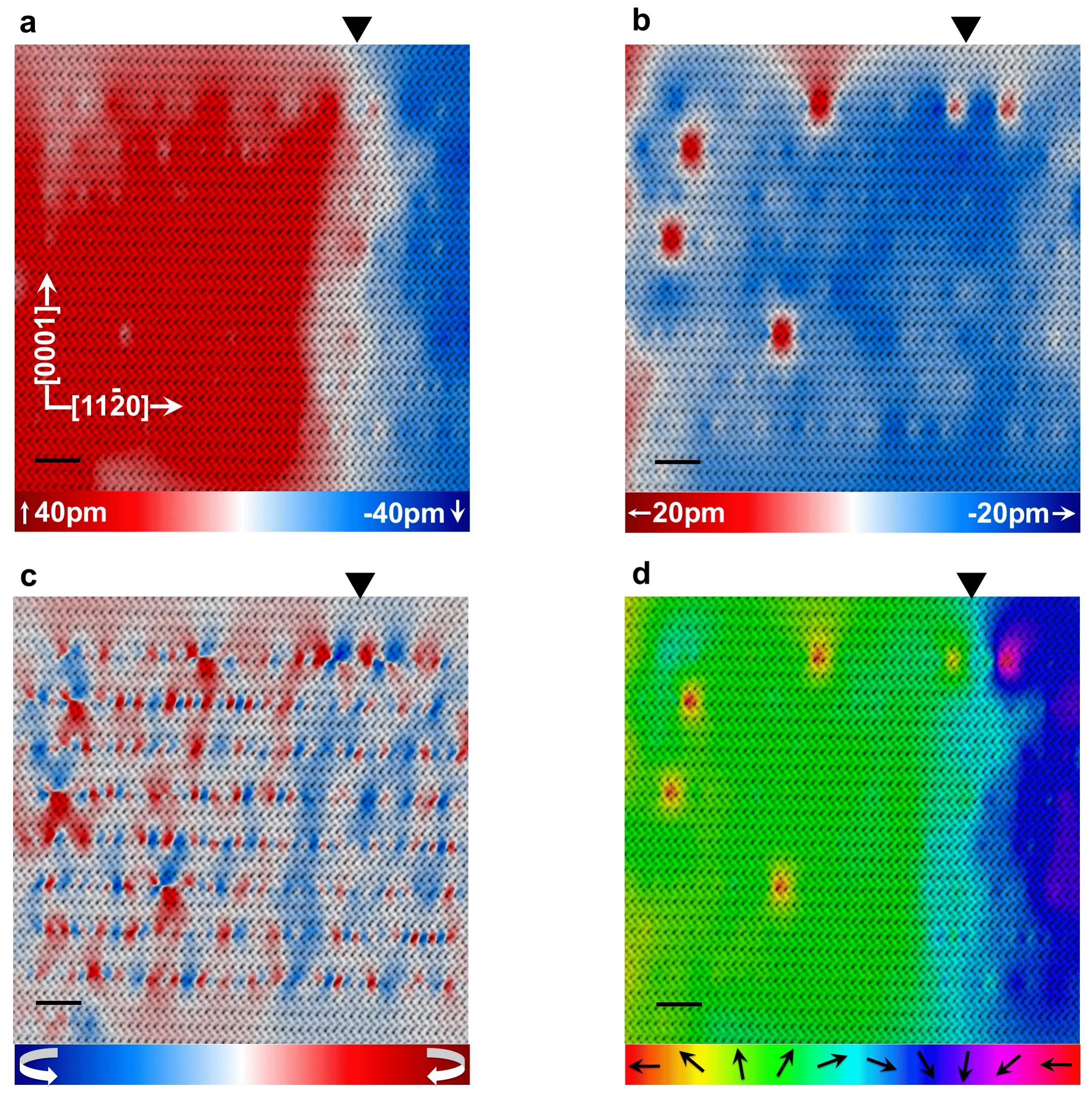}
	\caption{\label{fig:DW3} Domain wall in Region 3. (a) Ising displacements at region 3 of the domain wall with non-equivalent polarization on either sides. (b) N{\'e}el displacements demonstrating the presence of consistent and uniform N{\'e}el components in contrast to region 2 (\autoref{fig:DW2}). (c) Curl of the polar niobium-oxygen displacement map with a small discernible change at the domain wall. (d) Rotation map of the polar niobium-oxygen displacements. Scale bar in all images is 2nm.}
\end{figure}

\begin{figure}
	\centering
	\includegraphics[width=\columnwidth]{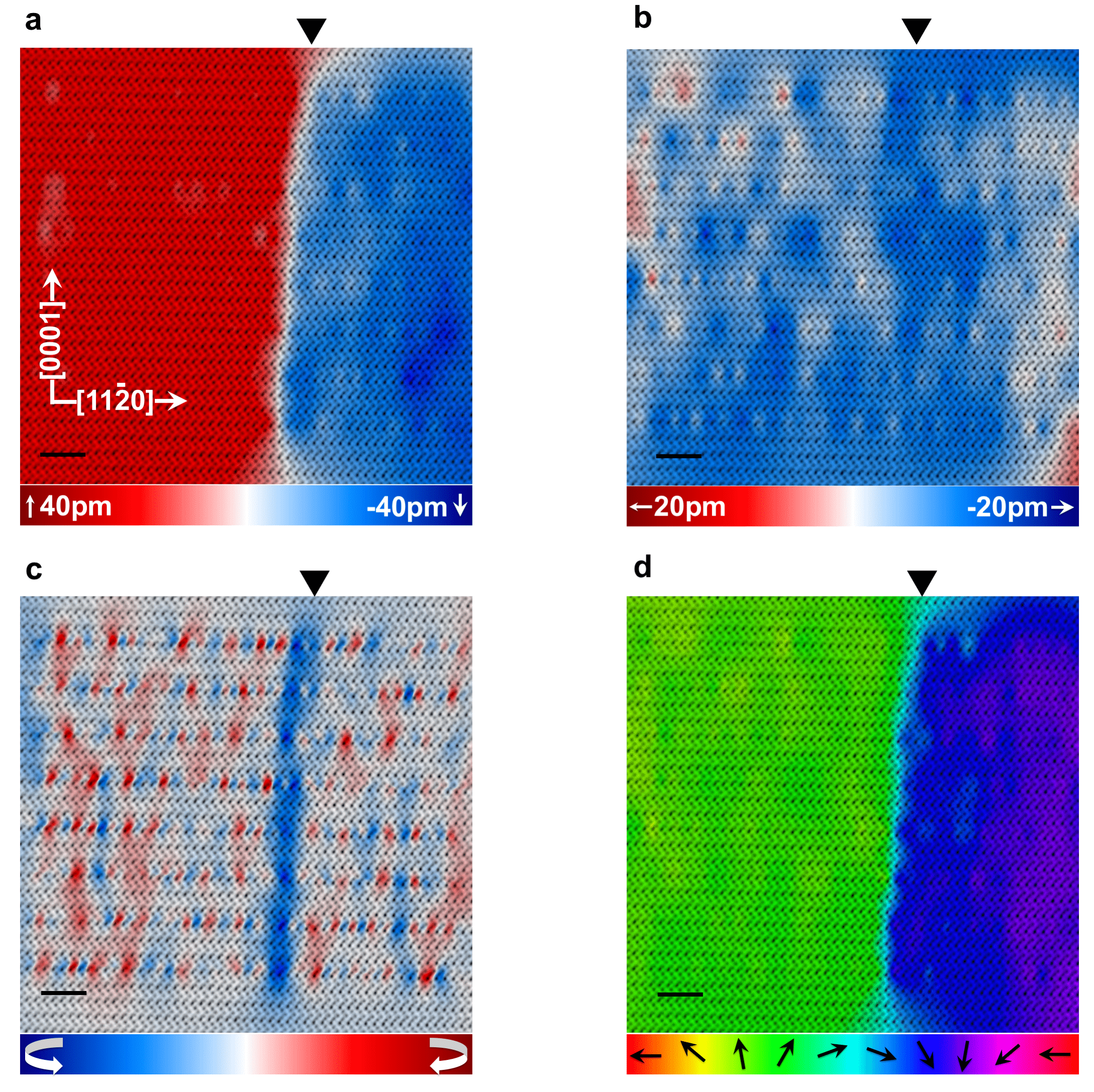}
	\caption{\label{fig:DW4} Domain wall in Region 4. (a) Ising displacements at region 4 of the domain wall with non-equivalent polarization on either sides. (b) N{\'e}el displacements demonstrating the presence N{\'e}el regions, not limited to only the domain wall. (c) Curl of the polar niobium-oxygen displacement map with a significant change only at the domain wall. (d) Rotation map of the polar niobium-oxygen displacements. Scale bar in all images is 2nm.}
\end{figure}

\begin{figure}
	\centering
	\includegraphics[width=\columnwidth]{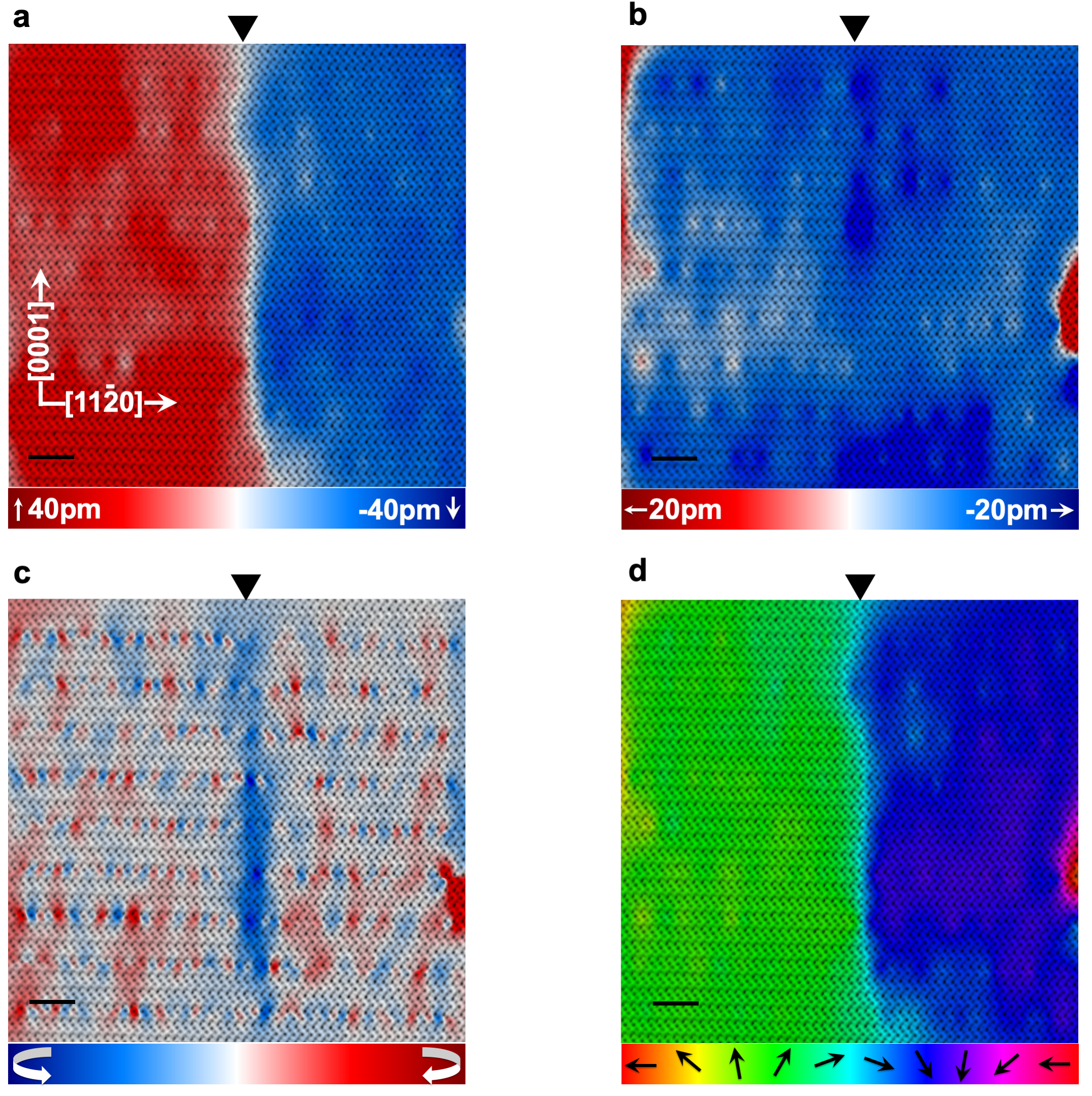}
	\caption{\label{fig:DW5} Domain wall in Region 5. (a) Ising displacements at region 5 of the domain wall with a kink in the wall. (b) Quasi uniform N{\'e}el displacements in the proximity of the domain wall. (c) Curl of the polar niobium-oxygen displacement map. (d) Rotation map of the polar niobium-oxygen displacements, with the Ising kink being visible. Scale bar in all images is 2nm.}
\end{figure}

Four other domain wall regions (labeled as regions 2-5) in addition to the region imaged in the main text (\autoref{fig:polarMap}) were imaged in the electron microscope, as demonstrated in \autoref{fig:DW2}, \autoref{fig:DW3}, \autoref{fig:DW4} and \autoref{fig:DW5}. As could be observed from all the systems the domain wall is consistently associated with significant N{\'e}el type non-Ising distortions. One of the regions of the domain wall, \autoref{fig:DW2} also demonstrates N{\'e}el distortions in both positive and negative directions, with leftward N{\'e}el distortions precipitating primarily at the domain wall. Also, the thickness of the Ising component at the domain wall is not uniform at different regions of the domain wall, with \autoref{fig:DW3} demonstrating significantly wider walls compared to the other regions imaged.

\section{\label{sec:ChargeCalc}Estimation of charge accumulation and electrostatic potential energy}

We roughly estimated the charge accumulation from the polar distortions to get an estimate in the energy magnitudes of electrostatic potential energy and the thermodynamic free energy decrease from the entropy. 
\begin{table*}
	\centering
	\begin{tabular}{llccc}
		\toprule
		\textbf{Atom} & \textbf{Cartesian Direction}	& \multicolumn{3}{c}{\textbf{Born Effective Charge}}\\
		\cmidrule(l){3-5}
		& & 1 & 2 & 3 \\
		\midrule
		\multirow{3}{*}{Li} & 1 & $1.150619$ & $-1.860690\times10^{-16}$ & $5.756686\times10^{-16}$ \\
		& 2 & $-1.364568\times10^{-16}$ & 1.150619 & $4.525044\times10^{-16}$ \\
		& 3 & $1.097943\times10^{-16}$ & $1.898622\times10^{-13}$ & -1.103018 \\
		\multirow{3}{*}{Nb} & 1 & 8.330707 & -2.061953 & $2.317528\times10^{-16}$ \\
		& 2 & 2.061953 & 8.330707 & $3.418674\times10^{-16}$ \\
		& 3	 & $-8.225340\times10^{-1}$ & $3.464213\times10^{-12}$ & 9.199131 \\
		\multirow{3}{*}{O1} & 1 & -3.848460 & -1.191682 & -2.153211 \\
		& 2 & -1.191682 & -2.472424 & -1.243157 \\
		& 3 & -2.048031 & -1.182431 & -3.434050 \\
		\multirow{3}{*}{O2} & 1 & -1.784406 & $-3.133223\times10^{-17}$ & $2.586606\times10^{-16}$ \\
		& 2 & $1.249054\times10^{-17}$ & -4.536477 & 2.486314 \\
		& 3 & $1.357373\times10^{-16}$ & 2.364863 & -3.434050 \\
		\multirow{3}{*}{O3} & 1 & -3.848460 & 1.191682 & 2.153211 \\
		& 2 & 1.191682 & -2.472424 & -1.243157 \\
		& 3 & 2.048031 & -1.182431 & -3.434050 \\
		\bottomrule
	\end{tabular}
	\caption{\label{tab:Born}Calculated Born effective charges in \ce{LiNbO3}}
\end{table*}

\begin{figure*}
	\centering
	\includegraphics[width=0.66\textwidth]{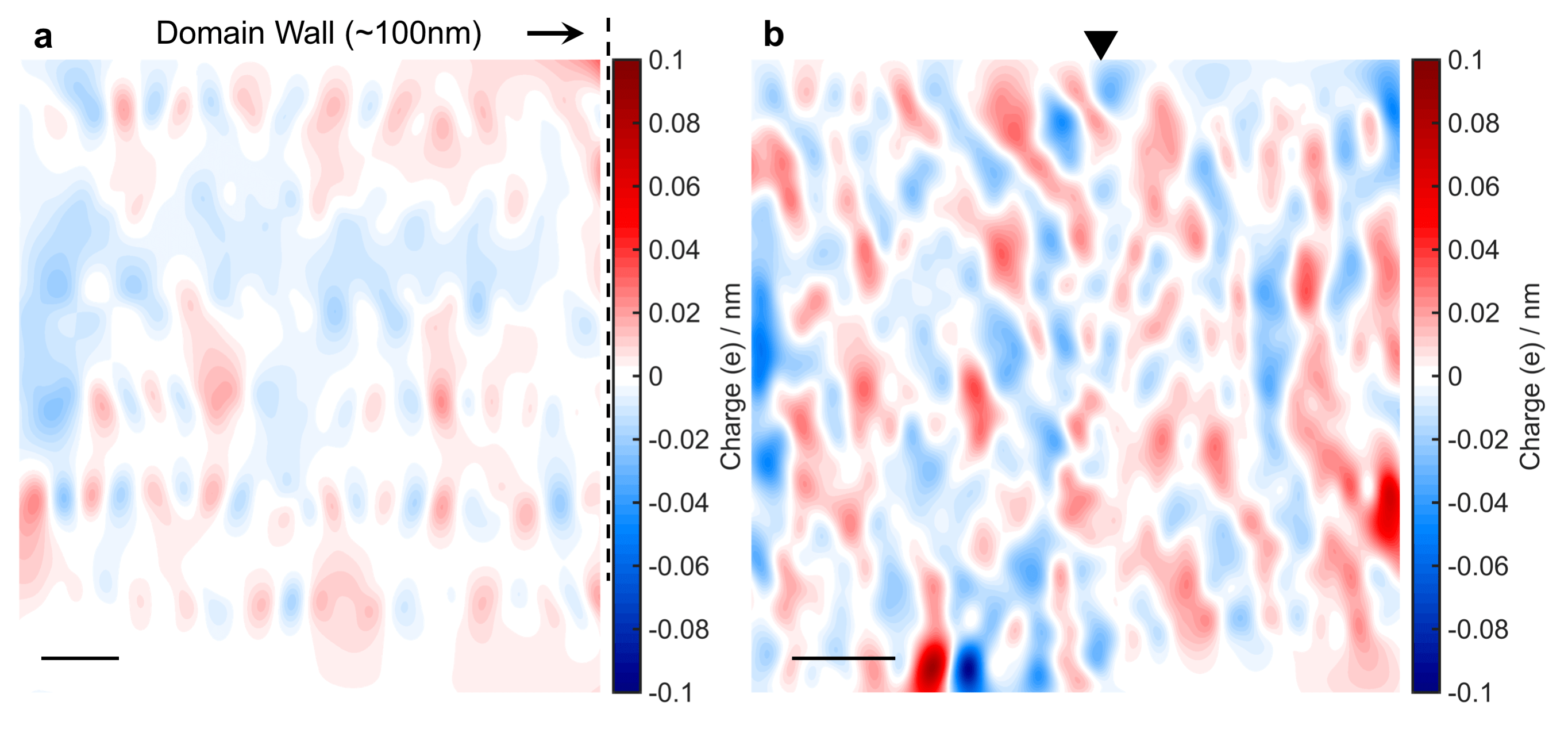}
	\caption{\label{fig:charge} Calculated charge accumulation. (a) Charge accumulation at a region of the bulk domain $\mathrm{\left( \approx 100nm \right) }$ away from the wall. The polarization maps are given in \autoref{fig:Domain1}(a) and  \autoref{fig:Domain1}(b) for the Ising and N{\'e}el  displacements respectively(b) Charge accumulation in the proximity of the domain wall (Region 1), with the black triangle showing the domain wall location. The polarization maps are given in \autoref{fig:polarMap}(a) and  \autoref{fig:polarMap}(b) for the Ising and N{\'e}el  displacements respectively. Scale bar in both images is 2nm.}
\end{figure*}

\begin{figure*}
	\centering
	\includegraphics[width=0.66\textwidth]{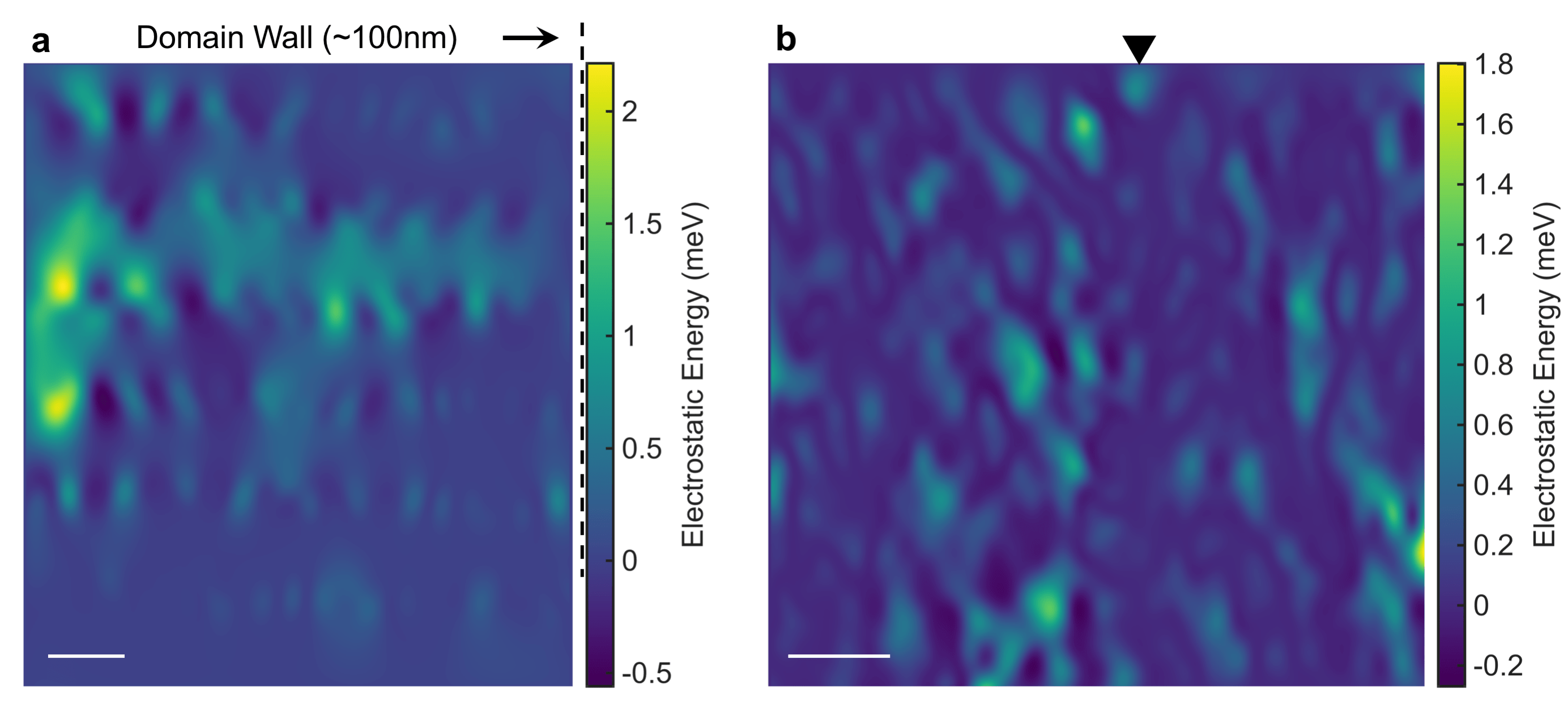}
	\caption{\label{fig:Electrostatic} Measured electrostatic potential energy. (a) Potential energy at a region of the bulk domain $\mathrm{\left( \approx 100nm \right) }$ away from the wall. The polarization maps are given in \autoref{fig:Domain1}(a) and  \autoref{fig:Domain1}(b) for the Ising and N{\'e}el  displacements respectively(b) Potential energy in the proximity of the domain wall (Region 1), with the domain wall location shown by the black triangle. The polarization maps are given in \autoref{fig:polarMap}(a) and \autoref{fig:polarMap}(b) for the Ising and N{\'e}el  displacements respectively. Scale bar in both images is 2nm. Potential energy was calculated with $\mathrm{\epsilon_r = 4.821}$\cite{LNO_dielectric}.}
\end{figure*}

Charge calculations were performed by first estimating the Born effective charge tensors theoretically, with the calculated Born effective charges presented in \autoref{tab:Born}. The calculated polar displacements from a representative bulk domain region (\autoref{fig:Domain1}) and a a representative domain wall region (\autoref{fig:polarMap}) respectively are vector multiplied with the Born effective charge tensors (\autoref{tab:Born}) for the niobium and oxygen atoms only, since we cannot image the lithium atoms. The divergence of this polarization is now the charge accumulation, which is presented in \autoref{fig:charge}, with \autoref{fig:charge}(a) showing the charge accumulation in the bulk domain region, and \autoref{fig:charge}(b) demonstrating the charge accumulation in the domain wall proximity. 

Thus, for each image, we have a total charge distribution. Assuming that each pixel corresponds to a charge value, then the total number of pixels (N) refers to the total possible charge values. The electrostatic potential energy is then calculated using \autoref{eq:ChargeEquation}, obtained through an integration of Coulomb’s law
\begin{equation}\label{eq:ChargeEquation}
\mathrm{
	U_E = \frac{1}{2} \Sigma_{x=1}^{N}q_x\Sigma_{y=1}^{N\left( y \neq x \right) } \left( \frac{1}{4\pi\epsilon_0\epsilon_{LiNbO_3}}\times\frac{q_y}{r_{xy}} \right)}
\end{equation}
where $q_x$ refers to the charge at a certain pixel, and $r_{xy}$ refers to the distance between distance between the $\mathrm{x^{th}}$ and the $\mathrm{y^{th}}$ pixel. The term $\sfrac{1}{2}$ prevents double counting the potential energy contribution between x and y, and y and x positions. The $\mathrm{\epsilon_{LiNbO_3}}$ is 4.821\cite{LNO_dielectric}. The calculated electrostatic potential energy for the two regions are shown in \autoref{fig:Electrostatic}(a) for the domain, and \autoref{fig:Electrostatic}(b) for the domain wall.

The electrostatic potential energy in the bulk domain from our calculations of polarization come out to be 0.37meV in the bulk domain and 0.45meV in the proximity of the domain wall. In contrast, the $-T\Delta S$ at 300K is -213meV at the bulk domain and -273meV in the proximity of the wall. Thus, the magnitudes are significantly different, and electrostatics would not prevent polar fluctuations.

\section{\label{sec:ImageSim}Simulation of \ce{LiNbO3} BF-STEM images}

\begin{table}
	\centering
	\begin{tabular}{ll}
		\toprule
		\textbf{Experimental Conditions}	& \textbf{Value}\\
		\midrule
		Crystal Structure	& \ce{LiNbO3} \\
		\multirow{3}{*}{Debye-Waller Parameters} & $\mathrm{u_{Li} = 0.67}${\AA} \\
		& $\mathrm{u_{Nb} = 0.3924}${\AA} \\
		& $\mathrm{u_O = 0.5}${\AA}\cite{LNO_XRD} \\ 
		\multirow{3}{*}{Lattice Parameters} & a = 5.172{\AA} \\
		& b = 5.172{\AA} \\ 
		& c = 13.867{\AA}\cite{LNO_neutron} \\
		Space Group & 161 (R3c)\cite{MegawLNO}\\
		Zone Axis	& $\mathrm{\left[ 1 \bar{1} 00 \right] }$\\
		Accelerating Voltage	& 200kV\\
		Inner Collection Angle	& 0mrad\\
		Outer Collection Angle	& 15mrad\\
		Cells & $\mathrm{1 \times 5}$\\
		Frozen Phonons & 10\\
		Slices per Unit Cell & 5\\
		Probe Semi-Angle	& 28mrad\\
		\bottomrule
	\end{tabular}
	\caption{\label{tab:SimSetup}BF-STEM simulation conditions in MacTempasX}
\end{table}

\begin{figure*}
	\centering
	\includegraphics[width=0.75\textwidth]{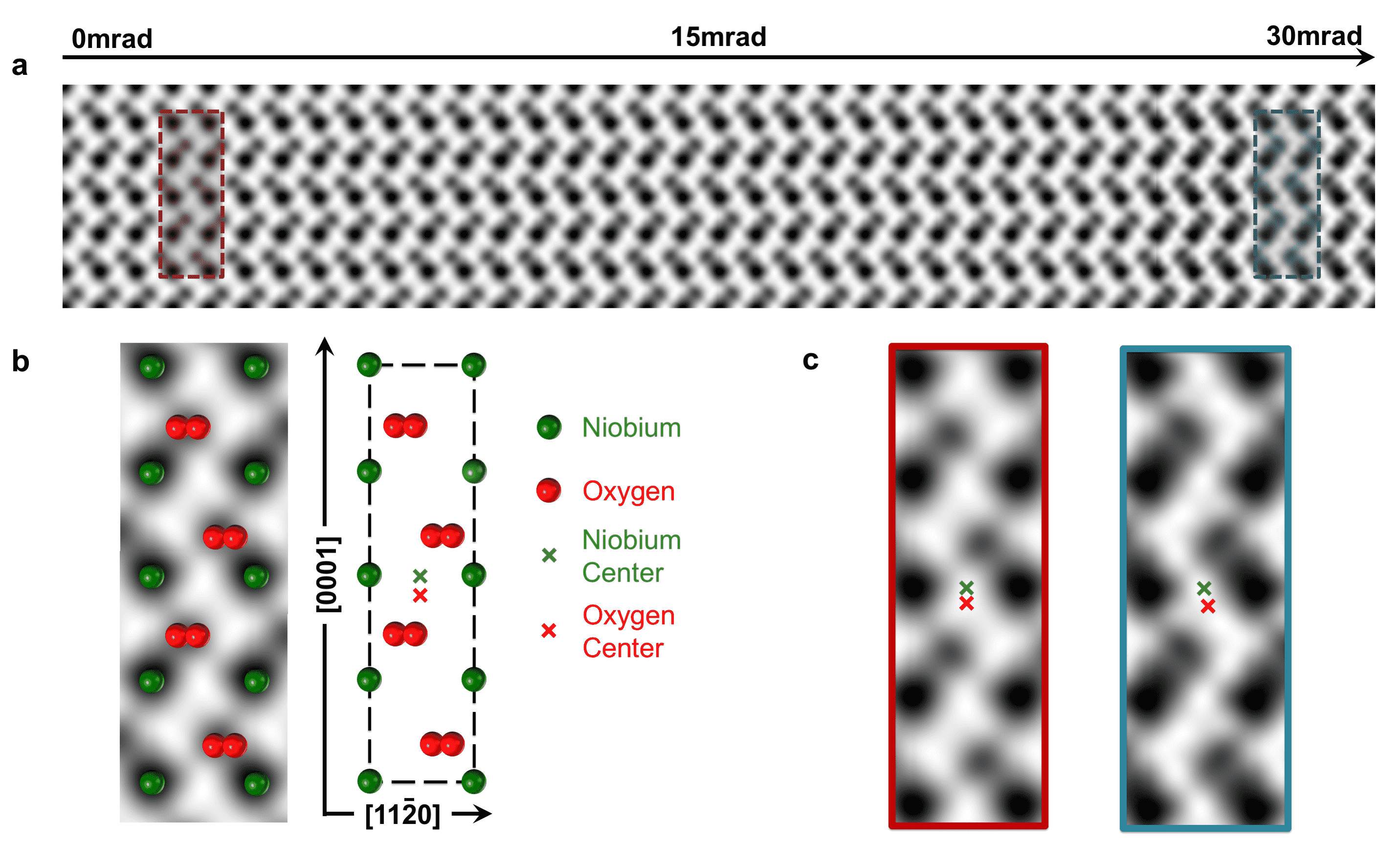}
	\caption{\label{fig:STEM_SIM} Evolution of BF-STEM image as a function of $\mathbf{\alpha}$ tilt. (a) Multislice simulations of BF-STEM image of \ce{LiNbO3} using the conditions detailed in \autoref{tab:SimSetup} without aberrations as a function of $\alpha$ tilt from 0 mrad (no tilt) to 30 mrad of $\alpha$ tilt(b) Zoomed in section with the niobium atoms in green and the oxygen atoms in red overlaid on top. (c) Comparison of the tilt effects at 0 mrad and 30 mrad showing how the relative displacement changes by 15pm in the $\mathrm{\left[ 000\bar{1} \right]}$ direction and by 25pm in the $\mathrm{\left[ 11\bar{2}0 \right]}$ direction.}
\end{figure*}

BF-STEM simulations of the \ce{LiNbO3} crystal structure were performed using the MacTempasX commercial software to understand the effect of tilt on imaging and atom position metrology, with the simulation parameters being enumerated in \autoref{tab:SimSetup}, with the effect of increasing $\alpha$ tilt being shown in \autoref{fig:STEM_SIM}\cite{mactempas}. As could be observed, the relative distance being the niobium-oxygen columns in sensitive to tilt, with the distance decreasing with increasing tilt. However, since the average Niobium-Oxygen polar Ising displacements match extraordinarily closely with the theoretical values in the domain wall figures presented in this work, tilt is not a contributing factor. Additionally, while increasing tilt would result in closer niobium-oxygen columns in the up domain, as shown in \autoref{fig:STEM_SIM}(c), it will also thus result in an increased distance in the down domain. However, the symmetric displacements observed (\autoref{fig:polarMap}, \autoref{fig:DW2}, \autoref{fig:DW3}, \autoref{fig:DW3}, \autoref{fig:DW4}, \autoref{fig:DW5}) would indicate this is not in fact the case. Additionally, the effects of tilts should be global, with a constant increase or decrease in the displacement measured over the entire field of view. This is however not the case in any of the images presented, with the changes in the Ising or N{\'e}el displacements occurring over only a few unit cells. Considering that \ce{LiNbO3} is a brittle oxide, and a 30mrad tilt would result in enormous local stresses, it is safe to assume that it is local displacements rather than tilt which is being observed here. 

\bibliographystyle{unsrtnat}
\renewcommand{\bibfont}{\small}
\bibliography{LNO_Entropy}
\end{document}